\documentclass[aps,pre,amsmath,amsfonts,amssymb,superscriptaddress,showpacs,11pt]{revtex4}
\usepackage{graphics}
\usepackage[latin1]{inputenc}
\usepackage{amscd}
\usepackage{soul}

\usepackage[pdftex]{graphicx}

\usepackage{rotating}
\usepackage[pdftex,hypertexnames=true,linktocpage=true]{hyperref}

\usepackage{color}

\usepackage{amsthm}
\usepackage{framed}
\usepackage{amssymb}
\usepackage{amsmath,mathrsfs}
\usepackage{hhline}
\usepackage{ulem}
\usepackage{fancyhdr}
\usepackage{tikz}
\everymath{\displaystyle}
\everymath{\displaystyle}
\everymath{\displaystyle}
\newcommand{\beq}{\begin{equation}}
\newcommand{\eeq}{\end{equation}}

\newcommand{\bi}{\begin{itemize}}
\newcommand{\ei}{\end{itemize}}
\newcommand{\D}{\mathrm{D}}

\newcommand{\bv}{\boldsymbol{v}}
\newcommand{\Hamilton}{\mathrm{H}}
\newcommand{\Fi}{\mathrm{F}}

\def\RR{\mathbb{R}}

\def\FF{\mathcal{F}}

\newcommand{\K}{\mathrm{K}}
\newcommand{\Tau}{\mathcal{T}}

\newcommand{\Ma}{\mathrm{M}}
\newcommand{\esse}{\mathcal{S}}
\newcommand{\bxi}{\boldsymbol{\xi}}
\newcommand{\bzeta}{\boldsymbol{\zeta}}

\newcommand{\Div}{\mathcal{D}}
\newcommand{\nihat}{\mathrm{X}}
\newcommand{\grad}{\mathrm{grad}}
\newcommand{\total}{\mathrm{d}}
\newcommand{\RC}{\mathrm{R}}

\newcommand{\Lc}{\overline{\nabla}_{\mbox{\small LC}}}

\newcommand{\bieta}{\boldsymbol{\eta}}
\newcommand{\Lagrange}{{\cal L}}

\newcommand{\metric}{{\rm g}}

\newcommand{\Riemann}{\mathcal{R}}
\newcommand{\tangent}{{\rm T}}

\newcommand{\paralleltransport}{{\rm P}}
\newcommand{\Lag}{{\rm L}}
\newcommand{\btheta}{\boldsymbol{ \theta}}
\newcommand{\E}{\mathrm{E}}
\newcommand{\action}{\mathrm{S}}

\newtheorem{definition}{Definition}[section]

\newtheorem{theorem}{Theorem}[section]    
   
\newtheorem{pro}{Proposition}[section]
    
\newtheorem{remark}{Remark}[section]

\begin{document}

\title{\textbf{Dynamical Systems induced by Canonical Divergence in dually flat manifolds}}
\author{Domenico Felice}
\email{felice@mis.mpg.de}
\affiliation{Max Planck Institute for Mathematics in the Sciences\\
 Inselstrasse 22--04103 Leipzig,
 Germany}
\author{Nihat Ay}
\email{nay@mis.mpg.de}
\affiliation{ Max Planck Institute for Mathematics in the Sciences\\
 Inselstrasse 22--04103 Leipzig,
 Germany\\
Santa Fe Institute, Santa Fe, NM 87501, USA}

\begin{abstract}
The principles of classical mechanics have shown that the inertial quality of mass is characterized by the kinetic energy. This, in turn, establishes the connection between geometry and mechanics. We aim to exploit such a fundamental principle for information geometry entering the realm of mechanics. According to the modification of curve energy stated by Amari and Nagaoka for a smooth manifold $\Ma$ endowed with a dual structure $(\metric,\nabla,\nabla^*)$, we consider $\nabla$ and $\nabla^*$ kinetic energies. Then, we prove that a recently introduced canonical divergence and its dual function coincide with  Hamilton principal functions associated with suitable Lagrangian functions when $(\Ma,\metric,\nabla,\nabla^*)$ is dually flat. Corresponding dynamical systems are studied and the tangent dynamics is outlined in terms of the Riemannian gradient of the canonical divergence. Solutions of such dynamics are proved to be $\nabla$ and $\nabla^*$ geodesics connecting any two points sufficiently close to each other. Application to the standard Gaussian model is also investigated.
\end{abstract}


\maketitle

\section{Introduction}\label{Intro}

\subsection{The role of geometry in classical mechanics}

The Riemannian geometry of a couple $(\Ma,\metric)$ is based on one single differential quantity called the {\it line element} $\total s$. This is defined in terms of the metric tensor $\metric$ by the following expression,
\begin{equation}
\label{lineelement}
\total s^2=\sum_{i,j=1}^n\ \metric_{ij}(\bxi)\ \total \xi^i\otimes\total \xi^j\ ,
\end{equation} 
and it  allows the development of a complete geometry on $\Ma$. Here, $\{\bxi\}$ is any system of local coordinates in the smooth manifold $\Ma$ and $n$ denotes the dimension of $\Ma$. In general, the coefficients $\metric_{ij}$ of the metric tensor are not constant over $\Ma$. However, when the geometry is Euclidean, the line element can be re-written by $\total s^2=\metric_{ij}\ \total \xi^i\otimes\total \xi^j$, with $\metric_{ij}=\delta_{ij}$ given by the Kronecker delta symbol and the Einstein's notation is adopted: from here on, whenever an index is repeated as sub and superscript in a product, it represents summation over the range of the index.

Geometry enters the realm of mechanics in connection with the {\it inertia of mass} which is characterized by the kinetic energy \cite{Lanczos}. The kinetic energy $\K$ of a single particle of mass $1$, is given in the $n$-dimensional Euclidean space by $\K=\frac{1}{2}\sum_{i=1}^n v_i^2$, where $\bv=(v_1,\ldots,v_n)$ is the velocity of the particle defined by
$$
\|\bv\|^2=\frac{(\total \xi^1)^2+\ldots+(\total \xi^n)^2}{\total t^2}\ ,
$$
and $\bxi=(\xi^1,\ldots,\xi^n)$ denotes the coordinates of the particle. Therefore, the line element in the $n$-dimensional space is defined by the following relation
$$
\total s^2=2\K\total t^2
$$
which establishes the connection between geometry and mechanics. The inertial quality of mass is expressed on the left-hand side of Newton's law in the form of {\it mass times acceleration} \cite{Lanczos}. The first law of dynamics states that if the vector sum of forces acting on the particle is zero, then the velocity of the particle is constant \cite{Arnold}. If this is the case, the trajectory of the particle of mass $1$ with velocity $\bv=\bxi_q-\bxi_p$ is the straight line
$$
\bxi(t)=\bxi_p+t \bv\ ,\qquad t\in [0,1]
$$  
where $\bxi_p$ and $\bxi_q$ are the local coordinates at $p$ and $q$, respectively, for any $p,q\in\Ma$.

\vspace{.2cm}

The extension of these concepts to the general Riemannian geometry can be expressed through the Lagrangian formulation of mechanics \cite{Lanczos} by requiring some {\it kinematical conditions} between the coordinates. We may then restrict the free movability of the particle by forcing it to stay along an arbitrary curve $\gamma(t)$. Hence, the kinetic energy along $\gamma(t)\ (t\in[0,1])$ is defined as
\begin{equation}
\label{kinetic}
\K=\Lagrange_{\metric}:= \frac{1}{2}\langle\dot{\gamma}(t), \dot{\gamma}(t)\rangle_{\gamma(t)}= \frac{1}{2}\|\dot{\gamma}(t)\|^2_{\gamma(t)}\ ,
\end{equation}
where $\langle\cdot,\cdot\rangle_{\gamma(t)}$ denotes the inner product and $\|\cdot\|_{\gamma(t)}$ is the norm, both induced by $\metric$ on $\gamma(t)$. The Hamilton's principle requires that the time-integral of the {\it Lagrangian function} $\Lagrange_{\metric}$ shall be stationary. Therefore, by defining the {\it energy of the curve} $\gamma$ as
\begin{equation}
\label{energycurveR}
\E(\gamma):=\int_0^1 \Lagrange_{\metric}(\gamma(t),\dot{\gamma}(t))\ \total t\ ,
\end{equation}
the Hamilton's principle is formulated by $\delta \E(\gamma)=0$, where $\delta\E$ is induced by infinitesimal variation $\delta\gamma$ of the trajectory under the constrains that $\delta \gamma(0)=\delta\gamma(1)=0$. Working with the local coordinates $(\xi^1(\gamma(t)),\ldots\xi^n(\gamma(t)))$, this principle, also called {\it the principle of the least action}, yields the Euler-Lagrange equations associated with $\Lagrange_{\metric}$ \cite{Taylor},
\begin{equation}
\label{E-Lequation}
\frac{\total}{\total t}\frac{\partial \Lagrange_{\metric}}{\partial \dot{\xi}^i}-\frac{\partial\Lagrange_{\metric}}{\partial \xi^i}=0,
\end{equation}
and it selects the definite path ``chosen by nature as the actual path of the motion" \cite{Lanczos}.  Then, the Euler-Lagrange equations for the energy $\E$ are 
\begin{equation}
\label{geodequationsR}
\ddot{\xi}^i(t)+\overline{\Gamma}^{i}_{jk}\dot{\xi}^i(t)\dot{\xi}^j(t)=0, \quad i=1,\ldots,n\ ,\quad \gamma(0)=p,\,\gamma(1)=q
\end{equation}
with $\overline{\Gamma}^{i}_{jk}$ denoting the Christoffel's symbols of the Levi-Civita connection $\Lc$ and  we used the abbreviation $\dot{\xi}^i(t)=\frac{\total}{\total t}\left(\xi^i(\gamma(t)\right)$ (see, e.g. \cite{Jost17}). The evaluation of the energy $\E(\gamma_c)$ at the $\Lc$-geodesic $\gamma_c$, i.e. the solution of Eq. \eqref{geodequationsR}, gives a two points function
\begin{equation}
\label{twopoints}
\action(p,q)=\E(\gamma_c)
\end{equation}
which is known in literature as the {\it Hamilton principal function} \cite{Taylor} associated with the Lagrangian $\Lagrange_{\metric}$. We will discover later in the paper the important role played by the Hamilton principal function for describing dynamical systems in a generalization of the Riemannian geometry. Therefore, given the Lagrangian $\Lagrange_{\metric}$ the Hamilton's principle asserts that the actual motion realized in nature is that particular motion for which the energy $\E$ assumes its smallest value \cite{Lanczos}. This principle is strengthened by the geometric interpretation of $\Lc$-geodesics. Indeed, a classical result in Riemannian geometry states that the path between $\gamma(0)=p$ and $\gamma(1)=q$ with minimum length is the $\Lc$-geodesic parametrized with respect to the length arc \cite{Lee97}. This establishes a close connection between mechanics and Riemannian geometry.

Information Geometry is a generalization of Riemannian geometry where not only Levi-Civita geodesics are considered. This poses the problem on interpreting more general geodesics in mechanistic terms. This paper is addressing this problem by introducing a generalization of the kinetic energy. The following section summarizes the main results of the paper.

\subsection{Outline of the main results}

\noindent In Information Geometry (IG) \cite{Amari00}, the geometry on a smooth manifold $\Ma$ is induced by a dual structure $(\metric,\nabla,\nabla^*)$,  where $\nabla$ and $\nabla^*$ are linear connections on the tangent bundle $\tangent\Ma$ such that
\begin{equation}
\label{dualstructure}
X\,\metric\left(Y,Z\right)=\metric\left(\nabla_X Y,Z\right)+\metric\left(Y,\nabla^*_X Z\right) \quad \, \forall \; X,Y,Z\in\Tau(\Ma)\ ,
\end{equation}
and $\Tau(\Ma)$ denotes the space of vector fields on $\Ma$. Hereafter, we assume that both the connections, $\nabla$ and $\nabla^*$, are torsion-free and we refer to $\esse=(\Ma,\metric,\nabla,\nabla^*)$ as {\it statistical manifold}  \cite{Ay17}. The complete information of the geometric structure of $\esse$ is encoded in a distance-like function $\Div:\Ma\times\Ma\rightarrow\RR^+$ with $ \Div(p,q)=0\ \mbox{iff}\ p=q$. Such $\Div$ is usually referred to as {\it divergence function} on $\Ma$ and allows to recover the dual structure $(\metric,\nabla,\nabla^*)$ in the following way \cite{eguchi1992}:
\begin{align}
\label{metricfromdiv}
&\metric_{ij}(p)=-\left.\partial_i\partial_j^{\prime} \Div(\bxi_p,\bxi_q)\right|_{p=q},\\
&\Gamma_{ijk}(p)=-\left.\partial_i\partial_j\partial_k^{\prime} \Div(\bxi_p,\bxi_q)\right|_{p=q}, \; {\Gamma}^*_{ijk}(p)=-\left.\partial^{\prime}_i\partial^{\prime}_j\partial_k \Div(\bxi_p,\bxi_q)\right|_{p=q} ,\nonumber
\end{align}
where $\Gamma_{ijk}=\metric\left(\nabla_{\partial_i}\partial_j,\partial_k\right)$, ${\Gamma}^*_{ijk}=\metric\left(\nabla^*_{\partial_i}\partial_j,\partial_k\right)$ are the symbols of the dual connections $\nabla$ and $\nabla^*$, respectively. Here, 
$
\partial_i=\frac{\partial}{\partial \xi_p^i} \, , \partial^{\prime}_i=\frac{\partial}{\partial \xi_q^i}
$
and $\{\bxi_p:=(\xi_p^1,\ldots,\xi_p^n)\}$ and $\{\bxi_q:=(\xi_q^1,\ldots,\xi_q^n)\}$ are local coordinate systems at $p$ and $q$, respectively.

The concept of divergence function has been exploited in \cite{Amari00} to supply a definition of energy curve in IG. More precisely, let $(\Ma,\metric,\nabla,\nabla^*)$ be a statistical manifold, the energy of any arbitrary path $\gamma:[0,1]\rightarrow\Ma$ is given by 
\begin{equation}
\label{CanoicalDivgamma}
\D(\gamma):= \D_{\gamma}(\gamma(1),\gamma(0)) := \int \int_{0\leq s\leq t\leq 1} \metric_{\gamma}(s) \frac{\mu(t)}{\mu(s)}\, \total s\ \total t,\qquad \mu(s):= e^{\int_0^t \Gamma_{\gamma}(s)\,\total s}
\end{equation}
where $\metric_{\gamma}$ and $\Gamma_{\gamma}$ are the projection coefficients to $\gamma$ of $\metric$ and $\nabla$, respectively. The energy $\D(\gamma)$ is usually referred as the $(\metric,\nabla)$-{\it divergence of the curve $\gamma$} from $\gamma(1)$ to $\gamma(0)$. Very remarkably, $\D(\gamma)$ does not depend on the parametrization $t\mapsto\gamma(t)$ of $\gamma$ but its orientation, and the $(\metric,\nabla^*)$-{\it divergence $\D^*(\gamma)$ of $\gamma$} coincides with the $(\metric,\nabla)$-divergence of the reversly oriented curve \cite{Amari00}. Hence, it turns out that $\D^*(\gamma)=\D_{\gamma}(\gamma(0),\gamma(1))$.

Recently, a divergence function has been proposed by using the geodesic integration of the inverse exponential map \cite{Ay15}. Such a divergence, named henceforth by {\it canonical divergence}, assumes the following expression,
\begin{equation}
\label{AyCD}
\Div(p,q)=\int_0^1\ t\,\langle\dot{\sigma}(t),\dot{\sigma}(t)\rangle_{\sigma(t)}\ \total t,
\end{equation} 
where $\sigma$ is the $\nabla$-geodesic connecting $p=\sigma(0)$ and $q=\sigma(1)$. In this manuscript, we show that for any arbitrary curve $\widetilde{\gamma}(\tau)$ we can consider the re-parametrized curve  $\gamma(t(\tau))=\widetilde{\gamma}(\tau)$ such that
$$
\D^*(\gamma)=\int_0^1t\ \|\dot{\gamma}(t)\|^2\ \total t=\Div_{\gamma}(\gamma(0),\gamma(1))\ .$$
This amounts to require that the acceleration of $\gamma$ has no tangential component and then the function
\begin{equation}
\label{Lagrangian}
\Lagrange(t,\gamma,\dot{\gamma})=\frac{t}{2}\, \langle\dot{\gamma}(t),\dot{\gamma}(t)\rangle_{\gamma(t)}
\end{equation}
can be understood as the kinetic energy of $\gamma$ in IG. In this way, according to Eq. \eqref{energycurveR}, $\Div_{\gamma}(\gamma(0),\gamma(1))$ turns out to be twice the energy of the curve $\gamma$. The target of this manuscript is to provide a close relationship between mechanics and geometry in IG by analogy with classical mechanics and Riemannian geometry. Specifically, we aim to exploit the canonical divergence \eqref{AyCD} as playing the key role of this connection. We mainly focus our investigation on {\it dually flat manifolds} (see Appendix \ref{Appendix} for more details on this class of statistical manifolds). Then we supply the Euler-Lagrange equations for the energy $\Div_{\gamma}$ and prove that the solution is given by an unparametrized $\nabla$-geodesic. Therefore, by noticing that unparametrized $\nabla$-geodesics are the integral curves of $\grad_{\gamma(t)}\Div_p(\cdot)$, we prove that $\Div(p,q)$ coincides with the  {\it Hamilton principal function} associated with the Lagrangian function $\Lagrange$. To be more precise, by defining the energy of the curve $\gamma$ in analogy with Eq. \eqref{energycurveR} as
$$
\Div_{\gamma}(p,q)=\int_0^1\Lagrange(t,\gamma(t),\dot{\gamma}(t))\,\total t,\quad \gamma(0)=p,\,\gamma(1)=q\ ,
$$
we succeed to prove that the canonical divergence $\Div$ is a two point function on $\Ma$ such that 
\begin{equation}
\label{Hamiltonfunction}
\Div(p,q)=\Div_{\gamma_c}(p,q)\ ,
\end{equation}
where $\gamma_c$ is the unparametrized $\nabla$-geodesic from $p$ to $q$. In \cite{Ciaglia17} the Hamilton principal function associated with a suitably chosen Lagrangian function is showed to be a potential function for the dual structure $(\metric,\nabla,\nabla^*)$ on a smooth manifold $\Ma$. In this manuscript we take the opposite avenue and prove that the canonical divergence proposed in \cite{Ay15}, which is a potential function for the dual structure $(\metric,\nabla,\nabla^*)$, turns out to be the Hamilton  principal function associated with the Lagrangian function \eqref{Lagrangian}. 

The relevance of the result stated by Eq. \eqref{Hamiltonfunction} can be evaluated in the Hamiltonian description of the mechanics \cite{Lanczos}. So, we address our investigation to the Hamiltonian formulation by defining a new function, i.e. the Hamiltonian of the system, through the Legendre transform \cite{Taylor}. More specifically, we define the conjugate momentum $\bzeta$ to $\bxi$ by means of the Lagrangian $\Lagrange$, i.e. $\bzeta=\partial\Lagrange/\partial\dot{\bxi}$, and then the Hamiltonian $\Hamilton$ associated with $\Lagrange$ is given by
$$
\Hamilton(t,\bxi(t),\bzeta(t)):=\zeta_i\,\xi^i-\Lagrange(t,\bxi(t),\dot{\bxi}(t))\ ,
$$
where $\bzeta=(\zeta_1,\ldots,\zeta_n)$. In this context, the equations of motion are given by the so-called Hamilton's equations for the dynamics,
\begin{equation}
\label{Heqs}
\dot{\xi}^i=\frac{\partial\Hamilton}{\partial\zeta_i},\quad \dot{\zeta}_i=-\frac{\partial\Hamilton}{\partial\xi^i},\qquad i=1,\ldots,n\ ,
\end{equation}
which substitute the Euler-Lagrange equations \eqref{E-Lequation}. Hamilton's equations consist of $2n$ first-order differential equations, while Lagrange's equations consist of $n$ second-order equations. However, Hamilton's equations usually do not reduce the difficulty of finding explicit solutions. The issue of finding transformations in the space of coordinates $\{(\bxi,\bzeta)\}$ which preserve the structure of Eq. \eqref{Heqs} and simplify it to a form in which the Hamilton's equations become directly integrable takes place in the Hamilton-Jacobi description of mechanics \cite{Lanczos}. The crucial point of this theory is that those transformations are completely characterized by knowing one single function $\action$, the generating function of the transformation. This function, commonly known as Hamilton principal function, is the solution of the Hamilton-Jacobi equation
$$
\Hamilton\left(t,\xi^i,\frac{\partial\,\action}{\partial\,\xi^i}\right)+\frac{\partial\,\action}{\partial\,t}=0,\quad i=1,\ldots,n
$$
where $\Hamilton$ is the Hamiltonian function \cite{Taylor}. This equation establishes a close connection between the Hamilton description of mechanics and the Hamilton-Jacobi theory. To see the relation between the Lagrangian formulation and the Hamilton-Jacobi theory, let firstly observe that in the most general case, $\action\equiv\action(\bxi,t)$ is a function of coordinates $\bxi$ and time $t$. Then, let assume that the transformation induced by $\action$,
$$
\zeta_i=\frac{\partial\,\action}{\partial\xi^i},\quad i=1,\ldots,n
$$
yields a solution of the equations of motion. By taking the total derivative of $\action$ and exploiting the Hamilton-Jacobi equation, we obtain
$$
\frac{\total\,\action}{\total\,t}=\sum_i\,\frac{\partial\,\action}{\partial\,\xi^i}\,\dot{\xi}^i+\frac{\partial\,\action}{\partial\,t}=\zeta_i\,\xi^i-\Hamilton=\Lagrange\ .
$$
Finally, by integrating with respect to $t$ we have that
$$
\action=\int\,\Lagrange(t,\bxi_c,\bzeta_c(\bxi,\dot{\bxi}))\,\total\,t\ ,
$$
where $(\bxi_c,\bzeta_c)$ is the solution of the equations of motion \cite{Taylor}. This proves that the Hamilton principal function is equal to the time integral of the Lagrangian function evaluated upon the solution of equations of motion.

In this article, we aim to employ the canonical divergence \eqref{AyCD} as Hamilton principal function for simplifying the Hamilton's equations of motion. In particular, we mainly focus on the equations expressed by $\dot{\xi}^i=\frac{\partial\Hamilton}{\partial\zeta_i},\,i=1,\ldots,n$ which are referred in literature to as the tangent dynamics \cite{Marmo15}. Thus, we show that these equations are differentiable gradient systems \cite{smale1967}. Finally, we succeed to prove that such dynamics is held by $\nabla$-geodesic. On the contrary, owing to the duality of the structure $(\metric,\nabla,\nabla^*)$ we obtain the same result for the dual $\Div^*$ of the canonical divergence \eqref{AyCD}. In this case, the trajectory of the tangent dynamics is given by the $\nabla^*$-geodesic.

As application of the theoretical approach so far described, we consider the standard Gaussian model. We then determine a nice physical interpretation of the tangent dynamics outlined in terms of the canonical divergence gradient system: the $\nabla$-geodesic corresponds to the well-known Uhlenbeck-Ornstein process which describes the probability that a free particle in
Brownian motion after a time $t$ has a velocity lying between $v$ and $v+\total v$, when it started at $t =0$ with the velocity $v_0$ \cite{ornstein}.

Usually, statistical manifolds are used to model families of probability distributions \cite{Ay17}. A pioneering work on the relationship between dynamical systems and probability distributions was established by Nakamura in \cite{Nakamura1993}. Here, gradient systems on the manifolds of Gaussian and multinomial distributions are shown to be completely integrable Hamiltonian systems. This result has been generalized by Fujiwara and Amari in \cite{Fujiwara95}. Here, the authors showed that the dualistic gradient flow can be characterized as completely integrable system and proved that solutions are unparametrized $\nabla$ and $\nabla^*$ geodesics. In the present article, we take a different approach. Specifically, by minimizing the energy curve $\Div_{\gamma}$ we may interpret the canonical divergence $\Div$ as the Hamilton principal function associated with the Lagrangian $\Lagrange$. Then, we exploit the properties of the Hamilton principal function in the Hamiltonian description of the mechanics and show that the tangent dynamics is a gradient system. Finally, we determine the solution of such a dynamics that is the $\nabla$-geodesic connecting any two points $p,q$ sufficiently close to each other.

The layout of the article is as follows. In Section \ref{KIG} we  show that the function $\Div_{\gamma}$ can be understood as modification of curve energy in Information Geometry. In Section \ref{GM} we prove that the canonical divergence $\Div$ is a Hamilton principal function associated with $\Lagrange$ and  we show that the tangent dynamics induced by $\Lagrange$ is given by the gradient flow induced by $\Div$. We then prove that dynamical paths between any two points sufficiently close to each other are $\nabla$ or $\nabla^*$ geodesics. Finally, we apply the methods so far described to the standard Gaussian model and compare them to the results obtained in \cite{Fujiwara95}. In Section \ref{Conclusions} we draw some conclusions by outlining the results obtained in this work and discussing possible extensions.  Finally, in Appendix \ref{Appendix} we present the basics of Information Geometry and we describe the canonical divergence  introduced in \cite{Ay15}.

\section{Kinetic energy and Information Geometry}\label{KIG}

Let $\esse=(\Ma,\metric,\nabla,\nabla^*)$ be a statistical manifold    and $\gamma:[a,b]\rightarrow\Ma\,(a<b)$ be a smooth curve in $\Ma$ connecting $p=\gamma(a)$ and $q=\gamma(b)$. The dual structure on $\gamma$ is then induced from $(\metric,\nabla,\nabla^*)$ by projection \cite{Amari00}. Therefore, the coefficients of $\metric_{\gamma}$, $\nabla_{\gamma}$ and $\nabla_{\gamma}^*$ corresponding to $\metric$, $\nabla$ and $\nabla^*$ are given by \cite{Amari00},
\begin{align}
\label{metricgamma}
& \metric_{\gamma}(t) = \metric_{ij}(\gamma(t))\, \dot{\gamma}^i\,\dot{\gamma}^j\\
\label{connectiongamma}
& \Gamma_{\gamma}(t) = \left\{ \dot{\gamma}^i(t)\,\dot{\gamma}^j(t) \Gamma_{kij}(t)+\ddot{\gamma}^j(t) \metric_{jk}(t)\right\}\dot{\gamma}^k(t) /\metric_{\gamma}(t)\\
\label{connectiongamma*}
& \Gamma^*_{\gamma}(t) = \left\{ \dot{\gamma}^i(t)\,\dot{\gamma}^j(t) \Gamma^*_{kij}(t)+\ddot{\gamma}^j(t) \metric_{jk}(t)\right\}\dot{\gamma}^k(t) /\metric_{\gamma}(t)\ .
\end{align}

Since $(\gamma,\metric_{\gamma},\nabla_{\gamma},\nabla_{\gamma}^*)$ is $1$-dimensional, it is dually flat. Now, Nagaoka and Amari introduced on dually flat manifolds a canonical divergence $\D$ of Bregman type (see Appendix \ref{Appendix}, Eq. \eqref{CanonicalDivergence}). In particular, this divergence is defined on $\gamma$. It assumes the following form \cite{Amari00},
\begin{equation}\label{generalDgamma}
\D(\gamma):= \D_{\gamma}(\gamma(b),\gamma(a)) = \int \int_{a\leq s\leq t\leq b} \metric_{\gamma}(s) \frac{\mu(t)}{\mu(s)}\, \total s\ \total t,\quad \mu(t):= e^{\int_a^t \Gamma_{\gamma}(s)\,\total s}
\end{equation}
and we call it \textit{$(\metric,\nabla)$-canonical divergence of the curve $\gamma$}. Very remarkably, $\D_{\gamma}$ does not depend on the parametrization $t\mapsto\gamma(t)$ but its orientation and the $(\metric,\nabla^*)$-divergence $\D^*$ of $\gamma$ coincides with the $(\metric,\nabla)$-divergence of the reversely oriented curve \cite{Amari00}. 

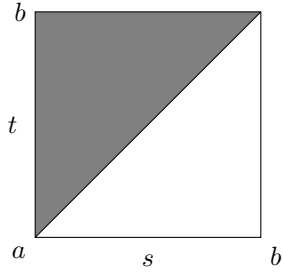
\begin{figure}
\centering
\begin{tikzpicture}
\draw (0,0) rectangle (3,3)
node at (1.5,-.3) {$s$}
node at (-.3,1.5) {$t$};
\draw (0,0) -- (3,3);
\coordinate [label=below left:$a$]
(a) at (0,0);
\coordinate [label=below right:$b$]
(a) at (3,0);
\coordinate [label= left:$b$]
(a) at (0,3);
\fill [black, opacity=0.5]
(0,0) -- (3,3) -- (0,3) -- (0,0);
\end{tikzpicture}
\caption{The domain of the definite integral in Eq. \eqref{generalDgamma} is highlighted in grey color.}\label{Region}
\end{figure}

\begin{remark}
\label{RDiv}
The statistical manifold $\esse=(\Ma,\metric,\nabla,\nabla^*)$ is {\it self-dual} when $\nabla=\nabla^*$. In this case, it turns out to be a Riemannian manifold endowed with the Levi-Civita connection $\Lc=\frac{1}{2}\left(\nabla+\nabla^*\right)$. Therefore, the projection of $\Lc$ to an arbitrary curve $\gamma:[a,b]\rightarrow\Ma$ is given by $\overline{\nabla}_{\gamma}=\overline{\Gamma}_{\gamma}(t)\ \dot{\gamma}$, where
$$
\overline{\Gamma}_{\gamma}(t)=\frac{\langle\left(\Lc\right)_{\dot{\gamma}}\dot{\gamma},\dot{\gamma}\rangle_{\gamma(t)}}{\langle \dot{\gamma},\dot{\gamma}\rangle_{\gamma(t)}}=\frac{1}{2}\frac{\frac{\total}{\total t}\metric_{\gamma}(t)}{\metric_{\gamma}(t)}\ ,
$$
and we used the well-known equality $\langle\left(\Lc\right)_{\dot{\gamma}}\dot{\gamma},\dot{\gamma}\rangle_{\gamma(t)}=\frac{1}{2}\frac{\total}{\total t}\langle\dot{\gamma},\dot{\gamma}\rangle_{\gamma(t)}$ \cite{Lee97}. Let us observe that the factor $\frac{\mu(t)}{\mu(s)}$ in the integral of Eq. \eqref{generalDgamma} can be re-written as follows,
$$
\mu(s,t):=\frac{\mu(t)}{\mu(s)}=e^{\int_s^t\overline{\Gamma}_{\gamma}(\tau)\total\tau}\ .
$$
By plugging $\overline{\Gamma}_{\gamma}$ into the $\mu$ factor of Eq. \eqref{generalDgamma}, we obtain
$$
\mu(s,t)= e^{\int_s^t \overline{\Gamma}_{\gamma}(\tau)\total \tau}=e^{\ln\left(\frac{\sqrt{\metric_{\gamma}(t)}}{\sqrt{\metric_{\gamma}(s)}}\right)}
= \frac{\sqrt{\metric_{\gamma}(t)}}{\sqrt{\metric_{\gamma}(s)}}\ .
$$
Consequently, from Eq.\eqref{generalDgamma} we obtain
$$
\D(\gamma)=\int\int_{a\leq s\leq t\leq b} \sqrt{\metric_{\gamma}(s)} \sqrt{\metric_{\gamma}(t)}\ \total s\,\total t\ .
$$
We may now observe from Fig. \ref{Region} that the area of the integration domain in Eq. \eqref{generalDgamma} is one half the area of the rectangle $[a,b]\times [a,b]$. Therefore, by the symmetry properties of $\metric_{\gamma}$ we obtain that
$$
\D(\gamma)=\frac{1}{2}\left(\int_a^b\sqrt{\metric_{\gamma}(s)}\ \total s\right)\left(\int_a^b\sqrt{\metric_{\gamma}(t)}\ \total t\right)=\frac{1}{2}\left(\int_a^b\sqrt{\langle\dot{\gamma}(t),\dot{\gamma}(t)\rangle_{\gamma(t)}}\ \total t\right)^2
$$
which proves that $\D(\gamma)$ is one half the square of the length of $\gamma$. This suggests to consider $\D(\gamma)$ as a modification of energy curve within Information Geometry.
\end{remark}

\begin{remark}
\label{ParametrizationInvariance}
In order to show that $\D(\gamma)$ does not depend on the parametrization of the curve, let us define $\widetilde{\gamma}(\tau)=\gamma(t(\tau))$.  Then, we have that
$$
\dot{\gamma}=\frac{1}{t^{\prime}(\tau)}\dot{\widetilde{\gamma}},\quad \ddot{\gamma}=\frac{1}{(t^{\prime}(\tau))^2}\left(\ddot{\widetilde{\gamma}}-\frac{t^{\prime\prime}(\tau)}{t^{\prime}(\tau)}\dot{\widetilde{\gamma}}\right)\ .
$$
By plugging these expressions into Eq. \eqref{metricgamma} and Eq. \eqref{connectiongamma}, we obtain
\begin{eqnarray*}
\metric_{\gamma}(t(\tau))&=& \metric_{\widetilde{\gamma}}(\tau)\frac{1}{(\tau^{\prime}(t))^2}\\ \,
\Gamma_{{\gamma}}(t(\tau))&=&\left\{\dot{\widetilde{\gamma}}^i(\tau)\dot{\widetilde{\gamma}}^j(\tau)\Gamma_{ijk}(\widetilde{\gamma}(\tau))+\ddot{\widetilde{\gamma}}^j(\tau)\,\metric_{jk}(\widetilde{\gamma}(\tau))\right\}\frac{\dot{\widetilde{\gamma}}^k(\tau)}{t^{\prime}(\tau)\,\metric_{\widetilde{\gamma}}(\tau)}\\
&&-\frac{t^{\prime\prime}(\tau)}{(t^{\prime}(\tau))^2}\frac{\dot{\widetilde{\gamma}}^j\dot{\widetilde{\gamma}}^k\,\metric_{jk}(\widetilde{\gamma}(\tau))}{\metric_{\widetilde{\gamma}}(\tau)}\\
&=& \Gamma_{\widetilde{\gamma}}(\tau)\,\frac{1}{t^{\prime}(\tau)}-\frac{t^{\prime\prime}(\tau)}{(t^{\prime}(\tau))^2}\ .
\end{eqnarray*}
Consider now the factor $\frac{\mu(t)}{\mu(s)}$ in the integral of Eq. \eqref{generalDgamma}, which is given by,
\begin{eqnarray*}
\mu(s(\tau),t(\tau))=\frac{\mu(t(\tau))}{\mu(s(\tau))}&=&e^{\int_{s(\tau)}^{t(\tau)}\Gamma_{{\gamma}}(\omega(\tau))\total \omega}\ .
\end{eqnarray*}
Then, the integral $\int_{s(\tau)}^{t(\tau)}\Gamma_{{\gamma}}(\omega(\tau))\total \omega$ can be computed by
\begin{eqnarray*}
\int_{s(\tau)}^{t(\tau)}\Gamma_{{\gamma}}(\omega(\tau))\total \omega&=& \int_{s(\tau)}^{t(\tau)}\Gamma_{{\widetilde{\gamma}}}(\tau)\frac{1}{\omega^{\prime}(\tau)}\total \omega-\int_{s(\tau)}^{t(\tau)}\frac{\omega^{\prime\prime}(\tau)}{(\omega^{\prime}(\tau))^2}\total\omega\\
&=& \int_{\tau^{-1}(s)}^{\tau^{-1}(t)}\Gamma_{\widetilde{\gamma}}(\tau)\total\tau-\int_{\tau^{-1}(s)}^{\tau^{-1}(t)}\frac{\omega^{\prime\prime}(\tau)}{\omega^{\prime}(\tau)}\total\tau\ ,
\end{eqnarray*}
where we adopted the change of variable rule $\total\omega=\omega^{\prime}(\tau)\total\tau$. The second integral of the latter equality can be computed as follows,
$$
\int_{s^{-1}(\tau)}^{t^{-1}(\tau)}\frac{\omega^{\prime\prime}(\tau)}{\omega^{\prime}(\tau)}\total\tau=\left.\ln\left[\omega^{\prime}(\tau)\right]\right|_{s^{-1}(\tau)}^{t^{-1}(\tau)}=\ln\frac{s^{\prime}(\tau)}{t^{\prime}(\tau)}\ .
$$
By collecting all these results we then perform the last calculation
\begin{eqnarray*}
\D(\gamma)&=&\int \int_{a\leq s\leq t\leq b} \metric_{\gamma}(s) \frac{\mu(t)}{\mu(s)}\, \total s\ \total t\\
&=& \int \int_{a\leq s(\tau)\leq t(\tau)\leq b} \frac{\metric_{\widetilde{\gamma}}(\tau)}{(s^{\prime}(\tau))^2}\frac{s^{\prime}(\tau)}{t^{\prime}(\tau)}\frac{\widetilde{\mu}(t(\tau))}{\widetilde{\mu}(s(\tau))}\,t^{\prime}(\tau)\total\tau\,s^{\prime}(\tau)\total\tau\\
&=&\int \int_{\tau^{-1}(a)\leq \tau^{-1}(s)\leq \tau^{-1}(t)\leq \tau^{-1}(b)} \metric_{\widetilde{\gamma}}(\tau)\frac{\widetilde{\mu}(t(\tau))}{\widetilde{\mu}(s(\tau))}\total\tau\total\tau =\D(\widetilde{\gamma})\ ,
\end{eqnarray*}
where $\mu(t(\tau))=e^{\int_a^{\tau^{-1}(t)}\Gamma_{\widetilde{\gamma}}(\tau)\total\tau}$.
\end{remark}

\vspace{.5cm}

\noindent From here on, we assume that the statistical manifold $(\Ma,\metric,\nabla,\nabla^*)$ is dually flat (for more details see Appendix \ref{Appendix}). Consider an arbitrary curve $\gamma:[0,1]\rightarrow\Ma$ such that $\gamma(0)=p$ and $\gamma(1)=q$. The intrinsic geometry $(\gamma,\metric_{\gamma},\nabla_{\gamma},\nabla_{\gamma}^*)$ of $\gamma$ is inherited from $(\metric,\nabla,\nabla^*)$ through the Eqs. \eqref{metricgamma}, \eqref{connectiongamma}, \eqref{connectiongamma*}. We aim to prove that, whenever $\gamma$ is $\nabla_{\gamma}$ or $\nabla_{\gamma}^*$ geodesic, the energy curve $\D^*(\gamma)$ reduces to the form
\begin{equation}
\label{reducedcurveenergy}
\Div_{\gamma}(p,q)=\int_0^1\,t\,\langle\dot{\gamma}(t),\dot{\gamma}(t)\rangle_{\gamma(t)}\,\total t\ 
\end{equation}
which is assumed by the canonical divergence introduced by Ay and Amari in \cite{Ay15} for an arbitrary curve $\gamma$.
This result amounts to require some kinematical conditions which pave the way for a nice interpretation of $\Div_{\gamma}$ as the time integral of twice the kinetic energy along $\gamma$. 
To be more precise, consider firstly the general form of $\D^*(\gamma)$,
\begin{equation}
\label{GeneralD*gamma}
\D^*(\gamma)=\int\int_{0\leq s\leq t\leq 1}\metric_{\gamma}(s)\ e^{\int_s^t \Gamma^*_{\gamma}(\tau)\total\tau}\total s \total t\ ,
\end{equation}
where $\Gamma_{\gamma}^*(t)$ is given by Eq. \eqref{connectiongamma*}. Since we assumed that $(\Ma,\metric,\nabla,\nabla^*)$ is dually flat, we can work with a $\nabla^*$-affine system of coordinates $\{\bxi^*\}$ and then write
\begin{equation}
\label{nabla*coord}
\gamma(t)=(\xi^{*1}(\gamma(t)),\ldots,\xi^{*n}(\gamma(t)))\ .
\end{equation}
Therefore, we can see from Eq. \eqref{connectiongamma*} that $\Gamma_{\gamma}^*$ can be written as follows:
\begin{equation}\label{Gamma*}
\Gamma_{\gamma}^*(t)=\frac{\langle\ddot{\gamma}(t),\dot{\gamma}(t)\rangle_{\gamma(t)}}{\langle\dot{\gamma}(t),\dot{\gamma}(t)\rangle_{\gamma(t)}}\ .
\end{equation}
We can now decompose the acceleration $\ddot{\gamma}$ by a tangential component along $\gamma$ plus a normal direction \cite{Lee97}:
\begin{equation}
\label{acceleration*}
\ddot{\gamma}=\left(\nabla^*_{\gamma}\right)_{\dot{\gamma}}\dot{\gamma}+II^*(\dot{\gamma},\dot{\gamma})\ ,
\end{equation}
where $II^*$ denotes the second fundamental form of $\gamma$ with respect to $\nabla^*$. Owing to the intrinsic geometry of $\gamma$, we can find a re-parametrization $\tau_{\nabla^*_{\gamma}}(t)$ such that $\gamma(\tau_{\nabla^*_{\gamma}}(t))$ is the $\nabla_{\gamma}^*$-geodesic from $p$ to $q$. Hence, we have that $\left(\nabla_{\gamma}^*\right)_{\dot{\gamma}}\dot{\gamma}=0$. From Eq. \eqref{acceleration*}, we can see then that $\ddot{\gamma}$ is normal to the trajectory of $\gamma$. Therefore, from Eq. \eqref{Gamma*} we have that $\Gamma^*_{\gamma}\equiv 0$ and consequently $\D^*(\gamma)$ assumes the following form,
\begin{eqnarray*}
\D^*(\gamma)&=&\int_0^1\, t \,\metric_{\gamma}(\tau_{\nabla_{\gamma}^*}(t))\,\total t\\
&=& \int_0^1\,t\,\langle\frac{\total(\gamma\circ\tau_{\nabla^*_{\gamma}})}{\total t}(t),\frac{\total(\gamma\circ\tau_{\nabla^*_{\gamma}})}{\total t}(t)\rangle_{\gamma(t)}\,\total t\\
&=&\Div_{\gamma}(p,q)\ .
\end{eqnarray*}
This result suggests the following physical interpretation: on one side $\left(\nabla_{\gamma}^*\right)_{\dot{\gamma}}\dot{\gamma}=0$ says that the virtual particle of mass $1$ subjected to the acceleration \eqref{acceleration*} has constant velocity in the direction of $\dot{\gamma}$; on the other side, $\ddot{\gamma}=II^*(\dot{\gamma},\dot{\gamma})$ implies that the sum of all force fields acting on the virtual particle are orthogonal to the trajectory of $\gamma$. Hence, we may think of some {\it holonomic constraints} forcing the virtual particle as moving along the curve $\gamma$. In this way, we can say that $\Div_{\gamma}$ is the action functional \cite{Lanczos} in the ambient space $(\Ma,\metric,\nabla^*)$ measured by twice the following kinetic energy,
\begin{equation}
\label{LagFlat}
\Lagrange(\bxi^*,\dot{\bxi}^*,t)= \frac{t}{2}\, \dot{\bxi}^{*T}\, G(\bxi^*)\, \dot{\bxi}^*\ .
\end{equation}
Here $\bxi^{*T}=(\xi^{*1},\ldots,\xi^{*n})$ and $\dot{\bxi}^{*T}=(\dot{\xi}^{*1},\ldots,\dot{\xi}^{*n})$ are the generalized coordinates and the generalized velocities, respectively. Moreover, $G(\bxi^*)=\left(\metric_{ij}(\bxi^*)\right)_{i,j}$, where $\metric_{ij}$ are the components of the metric tensor $\metric$.

\vspace{.2cm}

\noindent On the other hand, we may also work with $\nabla$-affine coordinates $\{\bxi(t)\}$ and then write 
\begin{equation}
\label{nablacoord}
\gamma(t)=(\xi^{1}(\gamma(t)),\ldots,\xi^{n}(\gamma(t)))\ .
\end{equation}
In this case, the coefficient of $\nabla^*_{\gamma}$ obtained by projecting $\nabla^*$ to $\gamma$ is given by
$$
\Gamma_{\gamma}^*(t)=\frac{\langle\nabla_{\dot{\gamma}}^*\dot{\gamma},\dot{\gamma}\rangle_{\gamma(t)}}{\langle\dot{\gamma},\dot{\gamma}\rangle_{\gamma(t)}}\ .
$$
However, thanks to the duality \eqref{dualstructure} of the geometric structure $(\metric,\nabla,\nabla^*)$ we can rewrite $\Gamma_{\gamma}^*$ as follows,
$$
\Gamma_{\gamma}^*(t)=\frac{\frac{\total}{\total t}\langle\dot{\gamma},\dot{\gamma}\rangle_{\gamma(t)}}{\langle\dot{\gamma},\dot{\gamma}\rangle_{\gamma(t)}}-\frac{\langle\ddot{\gamma},\dot{\gamma}\rangle_{\gamma(t)}}{\langle\dot{\gamma},\dot{\gamma}\rangle_{\gamma(t)}}\ ,
$$
where we used $\nabla_{\dot{\gamma}}\dot{\gamma}=\ddot{\gamma}$ as $\gamma(t)$ is written in $\nabla$-affine coordinates.
Again, the acceleration of $\gamma$ can be given in terms of a tangential component and a normal direction to the trajectory of $\gamma$:
$$
\ddot{\gamma}=\left(\nabla_{\gamma}\right)_{\dot{\gamma}}\dot{\gamma}+II(\dot{\gamma},\dot{\gamma})\ ,
$$
where $II$ is the second fundamental form of $\gamma$ with respect to $\nabla$. In analogy to the previous case, we can choose a re-parametrization $\tau_{\nabla_{\gamma}}(t)$ such that $\gamma(\tau_{\nabla_{\gamma}}(t))$ is the $\nabla_{\gamma}$-geodesic from $p$ to $q$. Hence, we have that $\left(\nabla_{\gamma}\right)_{\dot{\gamma}}\dot{\gamma}=0$. This implies that $\Gamma_{\gamma}^*(t)=\frac{\frac{\total}{\total t}\langle\dot{\gamma},\dot{\gamma}\rangle_{\gamma(t)}}{\langle\dot{\gamma},\dot{\gamma}\rangle_{\gamma(t)}}$. Consequently, we have that
$$
\int_s^t\Gamma_{\gamma}^*(\tau)\total\tau=\int_s^t\frac{\frac{\total}{\total t}\langle\dot{\gamma},\dot{\gamma}\rangle_{\gamma(t)}}{\langle\dot{\gamma},\dot{\gamma}\rangle_{\gamma(t)}}\total\tau=\log\left(\frac{\metric_{\gamma}(t)}{\metric_{\gamma}(s)}\right)\ .
$$ 
By plugging it in Eq. \eqref{GeneralD*gamma} we then obtain that
\begin{eqnarray*}
\D^*(\gamma)&=&\int\int_{0\leq s\leq t\leq 1}\metric_{\gamma}(s)\ e^{\log\left(\frac{\metric_{\gamma}(t)}{\metric_{\gamma}(s)}\right)}\total s\total t\\
&=&\int_0^1\ t\ \metric_{\gamma}(t)\total t =\int_0^1\,t\,\langle\frac{\total(\gamma\circ\tau_{\nabla_{\gamma}})}{\total t}(t),\frac{\total(\gamma\circ\tau_{\nabla_{\gamma}})}{\total t}(t)\rangle_{\gamma(t)}\,\total t\\
&=&\Div_{\gamma}(p,q)\ .
\end{eqnarray*}
Therefore, we may interpret $\Div_{\gamma}$ as the action functional in the ambient space $(\Ma,\metric,\nabla)$ measured by twice the following kinetic energy, 
\begin{equation}
\label{Lag*}
\Lagrange^*(t,\bxi,\dot{\bxi})=\frac{t}{2}\langle\dot{\bxi}(t),\dot{\bxi}(t)\rangle_{\bxi(t)}\ ,
\end{equation}
where $\{\bxi(t)\}$ denotes a $\nabla$-affine coordinate system.

\vspace{.1cm}

\noindent To sum up, we have showed that the function $\Div_{\gamma}$  introduced in \cite{Ay15} may be interpreted as the energy of an arbitrary curve $\gamma$ in the framework of Information Geometry. In particular, it is the time integral of twice the kinetic energy $\Lagrange$ in the ambient space $(\Ma,\metric,\nabla^*)$ and the time integral of twice the kinetic energy $\Lagrange^*$ in the ambient space $(\Ma,\metric,\nabla)$. In the next section, we shall prove that the dynamics associated with $\Lagrange$ and $\Lagrange^*$ are unparametrized $\nabla$-geodesics and unparametrized $\nabla^*$-geodesics, respectively. 

\section{Dynamics in dually flat manifolds}\label{GM}

Let $(\Ma,\metric,\nabla,\nabla^*)$ be a dually flat statistical manifold and $\gamma:[0,1]\rightarrow\Ma$ an arbitrary path such that $\gamma(0)=p$ and $\gamma(1)=q$. So far, we have introduced two kinetic energies along $\gamma$; the one in the ambient space $(\Ma,\metric,\nabla^*)$ is given by
\begin{equation}
\label{K*}
\K^*=\frac{t}{2}\langle\frac{\total(\gamma\circ \tau_{\nabla^*_{\gamma}})}{\total t}(t),\frac{\total(\gamma\circ \tau_{\nabla^*_{\gamma}})}{\total t}(t)\rangle_{\gamma(t)}\ ,
\end{equation}
where $\tau_{\nabla_{\gamma}^*}(t)$ is a reparametrization of $t$ such that $\left(\nabla^*_{\gamma}\right)_{\dot{\widetilde{\gamma}}}\dot{\widetilde{\gamma}}=0$ with $\widetilde{\gamma}=\gamma\circ\tau_{\nabla^*_{\gamma}}$. The second kinetic energy is defined in the ambient space $(\Ma,\metric,\nabla)$ and it is given by
\begin{equation}
\label{K}
\K=\frac{t}{2}\langle\frac{\total(\gamma\circ \tau_{\nabla_{\gamma}})}{\total t}(t),\frac{\total(\gamma\circ \tau_{\nabla_{\gamma}})}{\total t}(t)\rangle_{\gamma(t)}\ ,
\end{equation}
where $\tau_{\nabla_{\gamma}}(t)$ is a reparametrization of $t$ such that $\left(\nabla_{\gamma}\right)_{\dot{\widehat{\gamma}}}\dot{\widehat{\gamma}}=0$ with $\widehat{\gamma}=\gamma\circ\tau_{\nabla_{\gamma}}$.

\vspace{.2cm}

Working with the $\nabla^*$-affine coordinates \eqref{nabla*coord} and the $\nabla$-affine coordinates \eqref{nablacoord}, we can exploit the tools of the Lagrangian formulation of mechanics \cite{Taylor} for supplying the Euler-Lagrange equations for the energy curve $\Div_{\gamma}(p,q)$ within both the ambient spaces, $(\Ma,\metric,\nabla^*)$ and $(\Ma,\metric,\nabla)$. 

\subsection{Lagrangian approach to mechanics in dually flat manifolds}

In this section we aim to prove that
$$
\Div(p,q)=\min\{\Div_{\gamma}(p,q)\ |\ \gamma:[0,1]\rightarrow\Ma,\, \gamma(0)=p,\,\gamma(1)=q\}
$$
in the ambient space $(\Ma,\metric,\nabla^*)$. In order to carry out this result, we rely on the following representation of the canonical divergence introduced in \cite{Ay15},
\begin{equation}
\label{AyCDvector}
\Div(p,q)=\int_0^1\ \langle\dot{\sigma}_t(1),\dot{\sigma}(t)\rangle_{\sigma(t)}\ \total t\ ,
\end{equation}
where $\sigma_{t}(s)\ (0\leq s\leq 1)$ is the $\nabla$-geodesic from $p$ to $\sigma(t)$. The importance of this representation relies on the statement claimed by Theorem \ref{Thgrad}, namely $\dot{\sigma}_{t}(1)=\grad_{t}\Div_p(\cdot)$ (see Section \ref{Agradientfields} of the Appendix \ref{Appendix} for more details).

\vspace{.1cm}

Before claiming the next result we need the following definition,
\begin{definition}
\label{unparametrizedgeodesic}
A curve $\gamma(t)=(\gamma^1(t),\ldots,\gamma^n(t))$ is an {\it unparametrized} $\nabla$-geodesic if under suitable parametrization $t(\tau)$ it obeys the following geodesic equation
\begin{equation}
\label{nablageod}
\ddot{\gamma}^h(t(\tau))+\Gamma_{ij}^h(\gamma(t(\tau)))\ \dot{\gamma}^i(t(\tau))\dot{\gamma}^j(t(\tau))=0\ .
\end{equation}
\end{definition}

We are now ready to prove that the canonical divergence $\Div$ is a Hamilton principal function associated with the Lagrangian function \eqref{LagFlat}, i.e. the time integral of $\Lagrange$ evaluated at the solution of the Euler-Lagrange equations. More precisely, we are going to show that the solution of the Euler-Lagrange equations associated with $\Lagrange$ is an unparametrized $\nabla$-geodesic $\bxi_c(t)$ from $p$ to $q$ such that $t\ \dot{\bxi}_c(t)=\dot{\sigma}(\tau(t))$ with $\sigma$ being the $\nabla$-geodesic from $p$ to $q$. Then, by recalling that the integral curves of $\grad_t\Div_p(\cdot)$ are $\nabla$-geodesic starting from $p$, we succeed to prove  the following result.

\begin{theorem}
\label{ThE-LFlat}
Let $(\Ma,\metric,\nabla,\nabla^*)$ be a dually flat manifold and $p,q\in\Ma$ sufficiently close to each other. Let $\Lagrange$ be the Lagrangian function \eqref{LagFlat} set up in the $(\Ma,\metric,\nabla^*)$ space.
Then, the canonical divergence $\Div$ coincides with the Hamilton principle function associated with $\Lagrange$. In particular, we have that
\begin{align}
\label{DivHam}
& \Div(p,q)=\int_{0}^1\ \Lagrange(t,\bxi_c,\dot{\bxi}_c)\ \total t\ ,
\end{align}
where $\bxi_c$ is an unparametrized $\nabla$-geodesic from $p$ to $q$.
\end{theorem}
\noindent {\bf Proof.}\, In order to prove our claim, we need to solve the Euler-Lagrange equations of $\Lagrange$. For this reason, we firstly write these equations in the local $\nabla^*$-affine coordinates \eqref{nabla*coord} (for the reason of readability, we drop out the symbol ``$*$"). Therefore, the Lagrangian function \eqref{LagFlat} reads as follows,
$$
\Lagrange(t,\bxi,\dot{\bxi})=\frac{t}{2}\, \metric_{ij}\, \dot{\xi}^i\dot{\xi}^j\ .
$$
Then, by noticing that $\metric_{ij}$ depends only on $\bxi$ and not on $\dot{\bxi}$ we obtain
\begin{equation}
\label{Lagcoordinatesderivative}
\frac{\partial \Lagrange}{\partial \xi^k}=\frac{t}{2}\, \partial_k\, \metric_{ij}\,\dot{\xi}^i\dot{\xi}^j
\end{equation}
and
\begin{equation}
\label{Lagvelocitiesderivative}
\frac{\partial \Lagrange}{\partial \dot{\xi}^k}=\frac{t}{2}\,\left( \metric_{ik}\,\dot{\xi}^i+\metric_{kj}\,\dot{\xi}^j\right)\ .
\end{equation}
Then, it follows that
\begin{eqnarray}
\frac{\total}{\total t}\frac{\partial \Lagrange}{\partial \dot{\xi}^k} &=& \frac{\metric_{ik}\,\dot{\xi}^i+\metric_{jk}\,\dot{\xi}^j}{2}+\frac{t}{2}\,\left(\metric_{ik}\,\ddot{\xi}^i+\metric_{kj}\,\ddot{\xi}^j\right)+\frac{t}{2}\,\left(\partial_r\, \metric_{ik}\, \dot{\xi}^i\,\dot{\xi}^r+\partial_r\, \metric_{kj}\, \dot{\xi}^j\,\dot{\xi}^r\right)\nonumber\\
&=& t\, \metric_{ik}\,\ddot{\xi}^i+ \metric_{ik}\,\dot{\xi}^i+\frac{t}{2}\,\left(\partial_r\, \metric_{ik}\, \dot{\xi}^i\,\dot{\xi}^r+\partial_r\, \metric_{kj}\, \dot{\xi}^j\,\dot{\xi}^r\right),\label{totalderivativepartialLag}
\end{eqnarray}
where we used the symmetry $\metric_{ik}=\metric_{ki}$. Now, we can plug Eqs. \eqref{Lagcoordinatesderivative}, \eqref{totalderivativepartialLag} in Eq. \eqref{E-Lequation} and we arrive at
$$
t\,\left(2\,\metric_{ik}\,\ddot{\xi}^i+\partial_j\, \metric_{ik}\, \dot{\xi}^i\,\dot{\xi}^j+\partial_i\, \metric_{kj}\, \dot{\xi}^j\,\dot{\xi}^i-\partial_k\, \metric_{ij}\,\dot{\xi}^i\dot{\xi}^j\right)+2\,\metric_{ik}\,\dot{\xi}^i=0,
$$ 
where we changed $r$ to $j$ and $i$ in the second and third terms, respectively. Now, since $\bxi$ is $\nabla^*$-affine,  we have that $\partial_j\metric_{ik}=\Gamma_{jki}$. In addition this is $3$-symmetric tensor \cite{Amari00}. Therefore, multiplying by the inverse $\metric^{hk}$ of the metric tensor $\metric_{hk}$, we obtain
$$
\frac{t}{2}\,\left(2\ddot{\xi}^h+{\Gamma}_{ij}^h\,\dot{\xi}^i\,\dot{\xi}^j\right)+\dot{\xi}^h=0,\quad h=1,\ldots,n
$$
where the Christoffel symbols ${\Gamma}_{ij}^h$ of $\nabla$ are given by
$
{\Gamma}_{ij}^h=\metric^{hk}{\partial_k\, \metric_{ij}}
$
and the Einstein's notation is adopted. Finally, we get the Euler-Lagrange equations associated with $\Lagrange$,
\begin{equation}
\label{E-Leqs}
t\ \ddot{\xi}^h+\frac{t}{2} \Gamma_{ij}^h\dot{\xi}^i\dot{\xi}^j+\dot{\xi}^h=0\ , \quad h=1,\ldots,n\ .
\end{equation}

We may now observe that $t\ \ddot{\bxi}+\dot{\bxi}=\frac{\total }{\total t}\left(t\ \dot{\bxi}\right)$. In this way, Eq. \eqref{E-Leqs} becomes
$$
\frac{\total }{\total t} V^h+\Gamma^h_{ij}(\bxi(t))\ V_t^i\frac{\dot{\xi}^j}{2}=0\ ,
$$
where we defined $V(t):=t\ \dot{\bxi}(t)$. This implies that $V(t)$ is $\nabla$-parallel along the curve $\bxi(t)=:\widetilde{\bxi}(2t)$. By noticing that $V(t)=2t\dot{\widetilde{\bxi}}$ we can see that $V(t)$ is proportional to $\dot{\widetilde{\bxi}}$. Moreover, it is $\nabla$-parallel along $\widetilde{\bxi}$. Then we can conclude that $\bxi(t)$ is an unparametrized $\nabla$-geodesic from $p$ to $q$. Indeed, by choosing a re-parametrization $t(\tau)$ such that $\frac{\total}{\total \tau}(t(\tau))=2\ t$, in the new parametrization the velocity vector 
$$\dot{\sigma}(\tau)=2 t \dot{\widetilde{\bxi}}(t)=V(t) $$
is $\nabla$-parallel. To prove this, we can carry out the following calculation:
\begin{eqnarray*}
\frac{\total}{\total \tau}\dot{\sigma}^{ h}(\tau)&=& \frac{\total t(\tau)}{\total \tau} \frac{\total}{\total t}\left(2 t \dot{\widetilde{\xi}}^h\right)=2 t \frac{\total}{\total t}\left( t \dot{\xi}^h\right)\\
&=&2 t \left(\dot{\xi}^h(t)+\ddot{\xi}^h(t)\right)=2 t \left(-\frac{t}{2}\Gamma^h_{ij}\ \dot{\xi}^i\dot{\xi}^j\right)\\
&=& -\Gamma^h_{ij}(\sigma(\tau))\ \dot{\sigma}^{i}(\tau)\dot{\sigma}^{ j}(\tau)\ .
\end{eqnarray*} 
This proves that the curve $\sigma(\tau):=\widetilde{\bxi}(t(\tau))$ is the $\nabla$-geodesic from $p$ to $q$.

In order to conclude, let  $\bxi_c(t)$ be the unparametrized $\nabla$-geodesic which solves Eq. \eqref{E-Leqs}. Then, the curve $\sigma(\tau)=\bxi_c(t(\tau))$ is the $\nabla$-geodesic from $p$ to $q$. According to the result given by Theorem \ref{Thgrad} in the Appendix \ref{Appendix}, we know that the $\nabla$-geodesic $\sigma_{\tau}$ being part of the $\nabla$-geodesic $\sigma$ from $p$ to $q$ and corresponding to the interval $(\tau(0),\tau(t)]$ coincides with $\grad_t\Div_p(\bxi_c(t))$. This implies that $\dot{\sigma}=\frac{\grad_{\tau}\Div_p(\cdot)}{\tau}$. Therefore, by naming $\action(p,q)$ the evaluation of $\int_0^1\Lagrange(\bxi,\dot{\bxi})\total t$ at the critical path $\bxi_c(t)$, we obtain that

\begin{eqnarray*}
\action(p,q)&=&\int_0^1\ \Lagrange(\bxi_c,\dot{\bxi}_c)\ \total t=\int_0^1\  \langle t\dot{\bxi}_c,\frac{\dot{\bxi}_c}{2}\rangle_{\bxi_c(t)}\ \total t\\
&=& \int_0^1\  \langle \dot{\sigma}(\tau(t)),\dot{\widetilde{\bxi}}_c(t)\rangle_{\widetilde{\bxi}_c(t)}\ \total t = \int_0^1\  \langle \grad_{\tau(t)}\Div_p(\cdot),\frac{\dot{\widetilde{\bxi}}_c(t)}{\tau(t)}\rangle_{\sigma(\tau(t))}\ \total t\\
&=& \Div(p,q)\ .
\end{eqnarray*}
This proves claim  \eqref{DivHam}.  \hfill $\square$

\vspace{.3cm}

\noindent The dual function of $\Div$ has been also introduced in \cite{Ay15}. Likewise $\Div$, the dual canonical divergence $\Div^*$ can be viewed as the path integral of a specific vector field. More precisely, we have that
\begin{equation}
\label{AyCDvector*}
\Div^*(p,q)=\int_0^1\ \langle\dot{\sigma}^*_t(1),\dot{\sigma}^*(t)\rangle_{\sigma(t)}\ \total t\ ,
\end{equation}
where $\sigma^*_t(s)\ (0\leq s\leq 1)$ is the $\nabla^*$-geodesic from $p$ to $\sigma^*(t)$ and $\sigma^*$ is the $\nabla^*$-geodesic from $p$ to $q$. Again, the vector field $\dot{\sigma}^*_t(1)$ is a gradient vector field along $\sigma^*(t)$, i.e. $\dot{\sigma}_t^*(1)=\grad_t\Div^*_p(\cdot)$. This plays a key role for interpreting $\Div^*$ as the Hamilton principal function associated with the Lagrangian function \eqref{Lag*} in the $(\Ma,\metric,\nabla)$ ambient space. 

The Euler-Lagrange equations of $\Lagrange^*$ are given by
\begin{equation}
\label{E-Leqs*}
t\ \ddot{\xi}^h+\frac{t}{2} \Gamma_{ij}^{*h}\dot{\xi}^i\dot{\xi}^j+\dot{\xi}^h=0\ , \quad h=1,\ldots,n\ ,
\end{equation}
where $\Gamma_{ij}^{*h}$ are the Christoffel's symbols of the $\nabla^*$-connection. The solution of Eq. \eqref{E-Leqs*} is an unparametrized $\nabla^*$-geodesic $\bxi_c(t)$ such that $t\ \dot{\bxi}_c(t)=\dot{\sigma}^*(\tau)$ is the velocity vector of the $\nabla^*$-geodesic $\sigma^*$ from $p$ to $q$. Therefore, by applying the same methods of Theorem \ref{ThE-LFlat} we can prove that the dual canonical divergence $\Div^*$ given by Eq. \eqref{AyCDvector*} is a Hamilton principal function associated with the Lagrangian function $\Lagrange^*$. In particular, we obtain that
$$
\Div^*(p,q)=\min\{\Div_{\gamma}(p,q)\ |\ \gamma:[0,1]\rightarrow\Ma,\, \gamma(0)=p,\,\gamma(1)=q\}
$$
in the ambient space $(\Ma,\metric,\nabla)$.

\begin{remark}\label{Rem2}
By using the Lagrangian formalism, Theorem \ref{ThE-LFlat} yields $\D^*(\gamma)=\Div(p,q)$ when $\gamma$ is a $\nabla$-geodesic. This result is confirmed by the geometric viewpoint as claimed in \cite{Amari00}. Here, the authors showed  indeed that the $(\metric,\nabla)$-divergence of $\gamma$ coincides with the canonical divergence of Bregman type between $\gamma(1)$ and $\gamma(0)$ whenever $\gamma$ is either $\nabla$-geodesic or $\nabla^*$-geodesic.
\end{remark}

\subsection{Hamiltonian approach to mechanics in dually flat manifolds}

The Lagrangian formulation of mechanics \cite{Taylor} on the configuration manifold $\Ma$ is defined on the tangent bundle $\tangent\Ma$ in terms of the Lagrangian $\Lagrange:\tangent\Ma\rightarrow\RR$. Then, by considering a system of local coordinates $\bxi=\{\xi^1,\ldots,\xi^n\}$ on $\Ma$, we can define an action integral $\action$ which is a functional over the set of differentiable path $\bxi:[0,1]\rightarrow\Ma$ with fixed endpoints,
\begin{equation}
\label{action}
\action(\bxi)=\int_{0}^{1}\ \Lagrange\left(t,\bxi(t),\dot{\bxi}(t)\right)\ \total t\ .
\end{equation}
The evaluation of $\action$ at the solution of the Euler-Lagrange equations $\bxi_c$ gives a two-point function \cite{Ciaglia17},
\begin{equation}
\label{Hamiltonprincipal}
\action(p,q)=\action(\bxi_c), \quad p=\bxi(0) \, \mbox{and}\, q=\bxi(1)
\end{equation}
which is known in literature as the Hamilton principal function \cite{Taylor}. In the previous section, we proved that $\Div(p,q)=\int_0^1\Lagrange(t,\bxi_c,\dot{\bxi}_c)\total t$ when $\Lagrange$ is given by Eq. \eqref{LagFlat} and $\bxi_c$ is an unparametrized $\nabla$-geodesic from $p$ to $q$. In addition, we also proved that $\Div^*(p,q)=\int_0^1\Lagrange^*(t,\bxi_c,\dot{\bxi}_c)\total t$ when $\Lagrange^*$ is given by Eq. \eqref{Lag*} and $\bxi_c$ is an unparametrized $\nabla^*$-geodesic from $p$ to $q$.

The Hamilton principal function is the generating function of a canonical transformation in the phase space of the system \cite{Marmo15}. More precisely, the dynamics on the phase space is described by the Hamiltonian formulation of mechanics \cite{Taylor}. This is defined on the cotangent bundle $\tangent^*\Ma$ and it takes place by replacing the generalized velocity $\dot{\bxi}$ with the generalized momentum $\bzeta$. Specifically, the Legendre transform $\FF\Lag:\tangent\Ma\rightarrow\tangent^*\Ma$ relates the tangent bundle and the cotangent bundle as follows,
\begin{equation}
\label{momentum}
(\bxi,\dot{\bxi})\mapsto (\bxi^i,\bzeta)=\left(\bxi,\total_{\dot{\bxi}}	\Lagrange\right),\quad \total_{\dot{\bxi}}\Lagrange:\tangent\Ma\rightarrow\RR,\, 
\end{equation}
where $\total_{\dot{\bxi}}	\Lagrange$ denotes the differential of $\Lagrange$ at $\dot{\bxi}$. Owing to the metric tensor $\metric$ and the regularity of $\Lagrange$, the differential $\total_{\dot{\bxi}}	\Lagrange$ is given by
\begin{eqnarray*}
\total_{\dot{\bxi}}	\Lagrange(Y)&=&\langle \grad_{\dot{\bxi}}\Lagrange,Y\rangle_{\bxi(t)}=\langle\metric^{ij}\frac{\partial\Lagrange}{\partial\dot{\xi}^i}\partial_j,Y^k\partial_k\rangle_{\bxi(t)}\\
&=&\metric_{jk}\,\metric^{ij}\,\frac{\partial\Lagrange}{\partial\dot{\xi}^i}\,Y^k=\frac{\partial\Lagrange}{\partial\dot{\xi}^i}\,Y^i\ ,
\end{eqnarray*}
where we used the canonical identification $\tangent\tangent\Ma\simeq\tangent\Ma$ and the well-known expression $\grad\,f=\metric^{ij}\frac{\partial\,f}{\partial\xi^i}\partial_j$, with $\metric^{ij}$ the components of the inverse of $\metric$ and $\partial_i=\frac{\partial}{\partial\xi^i}$. The momentum conjugate to $\bxi$ is then given by
$$
\zeta_i=\frac{\partial\Lagrange}{\partial\dot{\xi}^i},\quad i=1,\ldots,n\ .
$$
In this way a Hamiltonian function can be defined by
\begin{equation}
\label{Hamiltonian}
\Hamilton:\tangent^*\Ma\rightarrow\RR,\quad \Hamilton(t,\bxi,\bzeta)=\zeta_i\,\dot{\xi}^i-\Lagrange(t,\bxi,\dot{\bxi}),
\end{equation}
where $\dot{\bxi}\equiv \dot{\bxi}(\bxi,\bzeta)$ is now function of $\bxi$ and $\bzeta$. Then, the Euler-Lagrange equations transform into Hamilton's equations,
\begin{equation}
\label{HamiltonEqs}
\frac{\total \xi^i}{\total t}=\frac{\partial \Hamilton}{\partial \zeta_i},\qquad \frac{\total \zeta_i}{\total t}=-\frac{\partial \Hamilton}{\partial \xi_i}\ .
\end{equation}

\vspace{.3cm}

\noindent The canonical transformation, i.e. the one that preserves the structure of Eq. \eqref{HamiltonEqs}, induced by the Hamilton principal function is given by 
\begin{equation}\label{actiomomentum}
\frac{\partial \action}{\partial\xi^i}=\zeta_i,\quad i=1,\ldots,n\ .
\end{equation}
Therefore, having in hands the canonical divergence $\Div$ which we proved is the Hamilton principal function associated with the Lagrangian function \eqref{LagFlat}, we can try to simplify the Hamilton equations and hopefully solve it in order to describe dynamical systems in a general dually flat statistical manifold. To this aim, let us perform the following computation
\begin{eqnarray}
\label{momentumflat}
\zeta_i&=&\frac{\partial\Lagrange(t,\bxi,\dot{\bxi})}{\partial \dot{\xi}^i}=\frac{\partial}{\partial \dot{\xi}^i}(\frac{t}{2}\,\metric_{ij}(\bxi)\,\dot{\xi}^i\,\dot{\xi}^j)\nonumber\\
&=& t\,\metric_{ij}(\bxi)\,\dot{\xi}^j\ .
\end{eqnarray}
Therefore, the fiber Legendre transform $\FF\Lag:\tangent\Ma\rightarrow\tangent^*\Ma$ and its inverse can be written as follows,
\begin{align}
\label{FiberLegendreflat}
& \bzeta^T=  t\, G(\bxi)\,\dot{\bxi}\\
& \dot{\bxi}= \frac{1}{t}\, G^{-1}(\bxi)\,\bzeta^T,
\end{align}
where $G(\bxi)=(\metric_{ij}(\bxi))_{i,j}$ and $G^{-1}$ denotes the inverse matrix of $G$. At this point, from Eq. \eqref{Hamiltonian} we obtain the Hamiltonian function of the system that reads as follows,
\begin{align}
\label{Hamiltonflat}
\Hamilton(t,\bxi,\bzeta) & =\frac{1}{2 t}\,\metric^{ij}(\bxi)\,\zeta_i\,\zeta_j \nonumber\\
&= \frac{1}{2 t}\, \bzeta\, G^{-1}(\bxi)\, \bzeta^T\ .
\end{align}

From Eq. \eqref{HamiltonEqs} we can compute the Hamilton equations of the dynamics,
\begin{align}
\label{HamiltonEqscoord}
\dot{\xi}^i=& \frac{1}{t} \, \metric^{ij}(\bxi)\,\zeta_j\\
\label{HamiltonEqsmomentum}
\dot{\zeta}_i =& \frac{1}{2 t} \, \partial_i \metric^{jk}(\bxi)\,\zeta_j\,\zeta_k\ .
\end{align}
Eq. \eqref{HamiltonEqscoord} is usually referred to as the tangent dynamics associated with the Lagrangian $\Lagrange$. The next result claims that such a dynamics can be understood in terms of gradient systems.

\begin{theorem}
\label{gradientflowThm}
Let $(\Ma,\metric,\nabla,\nabla^*)$ be a dually flat statistical manifold and consider the canonical divergence $\Div_p(q)$ as Hamilton characteristic function associated with $\Lagrange$. Then the equation
$
\dot{\xi}^i=\frac{\partial\,\Hamilton}{\partial\, \zeta_i}, \quad i=1,\ldots,n
$
becomes the following dynamical system
\begin{equation}
\label{gradientflowsol}
\dot{\bxi}(t)=\frac{1}{t}\, \grad_t \,\Div_p\ .
\end{equation}

\end{theorem}
\noindent {\bf Proof.}\,  According to Eq. \eqref{actiomomentum}, we have that $\zeta_j=\frac{\partial \Div_p }{\partial \xi^j}$. Then, from Eq. \eqref{HamiltonEqscoord} we obtain that
\begin{align*}
\dot{\xi}^i(t) & =\frac{1}{t}\ \metric^{ij} \zeta_j(t)= \frac{1}{t}\ \left(\grad_t \Div_p\right)^i,
\end{align*}
where we used  $\grad f=\metric^{ij}\, \frac{\partial_i f}{\partial\xi^i} \partial_j$, with $f$ any suitable smooth function and the sum over $i,j$ is intended. This proves Eq. \eqref{gradientflowsol}.
\hfill $\square$

\vspace{.2cm}

We are now ready to determine the trajectories of the tangent dynamics described by Eq. \eqref{gradientflowsol}. 

\begin{theorem}
\label{tangentdynamicsThm}
Let $(\Ma,\metric,\nabla,\nabla^*)$ be a dually flat statistical manifold and $p,q\in\Ma$ sufficiently close to each other. Then the tangent dynamics $\dot{\bxi}=\frac{1}{t}\grad_t\Div_p$ evolves along the $\nabla$-geodesic connecting $p$ and $q$.
\end{theorem}
\noindent {\bf Proof.}\,  We know that
$
\grad_t\Div_p=\dot{\sigma}_t(1)
$,
where $\sigma_t(s)$ is a $\nabla$-geodesic starting from $p$ (see Theorem \ref{Thgrad} in the Appendix \ref{Appendix}). Consider now the $\nabla$-geodesic $\sigma(t)$ from $p$ to $q$. Then, $\sigma_t(s)\ (0\leq s\leq 1)$ can be defined by $\sigma_t(s):=\sigma(s\ t)$. This implies that $\dot{\sigma}(t)=\frac{\dot{\sigma}_t(1)}{t}$ and then $\sigma(t)$ solves Eq. \eqref{gradientflowsol}.
 \hfill $\square$
 
\begin{remark}
By stepping back to the proof of Theorem \ref{ThE-LFlat}, we can see that the solution of the Euler-Lagrange equations associated with the Lagrangian function \eqref{LagFlat} is an unparametrized $\nabla$-geodesic from $p$ to $q$. In \cite{Fujiwara95}, the authors proved that the solution of the gradient system
$$
\dot{\eta}^i=-\metric^{ij}\,\partial_j \D(\bieta)
$$
converges to $q$ from $p$ along an unparametrized $\nabla$-geodesic. Here, $\bieta$ is the $\nabla^*$-affine coordinates of the point $p$ and $\D(\bieta)$ is the canonical divergence introduced by Amari and Nagaoka (see \eqref{CanonicalDivergence} in the Appendix \ref{Appendix}) from $q$ to $p(\btheta)$.

On the contrary, we obtained Eq. \eqref{gradientflowsol} by means of the Hamiltonian formalism. Thanks to this approach we succeed to prove that the solution of Eq. \eqref{gradientflowsol} is the $\nabla$-geodesic from $p$ to $q$.  
\end{remark}

\begin{remark}
Let $U\subset\Ma$ be an open set such that there exists an homeomorphism $\varphi:U\rightarrow\RR^n$ with $\varphi(p)=(\xi^i,\ldots,\xi^n)$ for all $p\in U$. A {\it gradient system} is a system of differential equations of the form \cite{smale1967}
$$
\dot{\bxi}=-\grad\, V(\bxi)\ ,
$$
where $V:\RR^n\rightarrow\RR$ is a $C^{\infty}$ function on $\varphi(U)$. The system of differential equations given by Eq. \eqref{HamiltonEqscoord} and Eq. \eqref{HamiltonEqsmomentum} is instead usually referred to as {\it Hamiltonian system} \cite{Marmo15}. This system defines the local flow of the Hamiltonian vector field $X_{\Hamilton}\in\Tau(\tangent^*\Ma)$ which is defined by
$$
X_{\Hamilton}=\left(\frac{\partial\Hamilton}{\partial\bzeta},-\frac{\partial\Hamilton}{\partial\bxi}\right)\ .
$$
The Hamilton equations can be put in the following form
$$
\left(\begin{array}{l}
\dot{\bxi}\\
\dot{\bzeta}
\end{array}\right)
=J\,\grad\,\Hamilton,\quad J=\left(\begin{array}{cc}
0&\mathrm{Id}_n\\
-\mathrm{Id}_n&0
\end{array}\right)
$$
where $J\,\grad\,\Hamilton=X_{\Hamilton}$ is known as the Hamilton symplectic gradient \cite{Marmo15}. This form of the Hamilton equations highlights a symplectic structure. Indeed, by the vector $X_{\Hamilton}$ we may define a symplectic form on $\tangent\tangent^*\Ma$ as
$$
\omega(X_{\Hamilton},Y):=\total\Hamilton(Y)=\sum_{i}\left(\frac{\partial\,\Hamilton}{\partial\xi^i}\frac{\partial\,Y}{\partial\zeta_i}-\frac{\partial\Hamilton}{\partial\zeta_i}\frac{\partial\,Y}{\partial\xi^i}\right)\ ,\quad Y\in\tangent\tangent^*\Ma\ .
$$

To sum up, a gradient system underlies a Riemannian structure, while a Hamiltonian system carries a symplectic structure. Our observation suggests that symplectic structures play important role also in Information Geometry (see related work \cite{Zhang17}.

Theorem \ref{gradientflowThm} selects one of the Hamilton equations, namely Eq. \eqref{HamiltonEqscoord}, and shows that it can be interpreted as gradient system as the canonical divergence $\Div$ is a function of the  coordinates $\{\bxi\}$.

\end{remark}

\vspace{.2cm}

The dual canonical divergence $\Div^*(p,q)$ is a Hamilton principal function associated with the Lagrangian \eqref{Lag*}. Therefore, we can apply the same arguments as above in order to obtain
a Hamiltonian function $\Hamilton^*$. In this case, the tangent dynamics is given by the following differential equation
\begin{equation}
\label{gradflow*}
\dot{\bxi}=\frac{1}{t}\ \grad_t \ \Div_p^*,
\end{equation}
where $\bxi$ is any path in $\Ma$ from $p$ to $q$ and $\Div_p^*$ is given by Eq. \eqref{AyCDvector*}. Finally, we succeed to determine the tangent trajectories of the dynamics \eqref{gradflow*} induced by the Lagrangian function $\Lagrange^*$.
\begin{theorem}
\label{tangentdynamicsThm*}
Let $(\Ma,\metric,\nabla,\nabla^*)$ be a dually flat statistical manifold and $p,q\in\Ma$. Then the tangent dynamics $\dot{\bxi}=\frac{1}{t}\grad_t\Div^*_p$ evolves along the $\nabla^*$-geodesic connecting $p$ and $q$.
\end{theorem}
\noindent {\bf Proof.}\, The dual version of Theorem \ref{Thgrad} (see Appendix \ref{Appendix}) holds true, as well. In particular the gradient of the divergence $\Div^*_p(\cdot)$ is given by $\nabla^*$-geodesics. Therefore, by using the same methods as in the proof of Theorem \ref{tangentdynamicsThm} we get the desired result. \hfill $\square$

\subsection{Example of Theorem \ref{tangentdynamicsThm}}

Let us consider the family of Gaussian probability distributions
with mean $\mu$ and variance $\sigma^2$:
\begin{equation}
\label{Gaussianmono}
p(x;\mu,\sigma)= \frac{1}{\sqrt{2 \pi}\,\sigma}\exp\left[-\frac{(x-\mu)^2}{2\sigma^2}\right],
\end{equation} 
where $\mu\in\RR$ and $\sigma\in\RR^+$ are the local coordinates of $\Ma=\{p\equiv p(x;\mu,\sigma)\}$. The information geometry of this statistical model is given in terms of the Fisher metric $\metric^{\Fi}$, whose components are computed by
\begin{equation}
\label{Fisher}
\metric_{ij}^{\Fi}=\int_{\RR}\, p(x;\mu,\sigma)\, \partial_i\ln\left(p(x;\mu,\sigma)\right)\partial_j\ln\left(p(x;\mu,\sigma)\right)\,\total x
\end{equation}
and in terms of the exponential connection $\nabla={\nabla}^e$ and the mixture connection $\nabla^*={\nabla}^m$. The connection symbols are given by the following relations,
\begin{align}
\label{exp&mixt}
& \Gamma_{ij,k}^e= \int_{\RR}\, p(x;\mu,\sigma)\, \partial_i\partial_j\ln\left(p(x;\mu,\sigma)\right)\partial_k\ln\left(p(x;\mu,\sigma)\right)\,\total x,\\
& \Gamma_{ij,k}^m= \int_{\RR}\, p(x;\mu,\sigma)\,\left( \partial_i\partial_j\ln\left(p(x;\mu,\sigma)\right)+\partial_i\ln\left(p(x;\mu,\sigma)\right)\partial_j\ln\left(p(x;\mu,\sigma)\right)\right)\partial_k\ln\left(p(x;\mu,\sigma)\right)\,\total x\ ,\nonumber
\end{align}
where $i,j\in\{1,2\}$ and $\partial_1=\frac{\partial}{\partial\mu}$ and 
$\partial_2=\frac{\partial}{\partial\sigma}$.

\vspace{.2cm}

Consider now two probability distributions of this family, namely $p\equiv p_1(x;\mu_1,\sigma_1)$ and $q\equiv p_2(x;\mu_2,\sigma_2)$. The tangent dynamics \eqref{gradientflowsol} can be obtained by the gradient system induced by the canonical divergence $\Div$. We now take this avenue for studying the tangent dynamics within the standard Gaussian model.  In order to compute $\Div(p_1,p_2)$ we exploit Eq. \eqref{AyCDvector}. In particular, we look for the $e$-geodesic $\sigma^e$ from $(\mu_1,\sigma_1)$ to $(\mu_2,\sigma_2)$ and the $1$-parameter family of $e$-geodesics $\sigma^e_t$ from $(\mu_1,\sigma_1)$ to $\sigma^e(t)=(\mu(t),\sigma(t))$.

According to Eq. \eqref{Fisher}, the Fisher metric of \eqref{Gaussianmono} is given by
\begin{equation}
\label{FisherGauss}
\metric^{\Fi}=\left(\begin{array}{cc}
\frac{1}{\sigma^2} & 0\\
0& \frac{2}{\sigma^2}
\end{array}\right)\ .
\end{equation}
Whereas, the non-zero connection symbols of $\nabla^e$ and $\nabla^m$ are obtained by Eq. \eqref{exp&mixt} and they read as follows,
\begin{align}
\label{exp&mixtSymbols}
& \Gamma_{121}^e=\Gamma_{211}^e=-\frac{2}{\sigma^3},\quad \Gamma^e_{222}=-\frac{6}{\sigma^3}\\
& \Gamma^m_{112}=\frac{2}{\sigma^3},\quad \Gamma^m_{222}=\frac{2}{\sigma^3}\ ,\nonumber
\end{align}
respectively. By denoting $\metric^{ij}$ the components of the inverse of $\metric^{\Fi}$, the Christoffel symbols are computed by $\Gamma_{ij}^k=\metric^{ij}\Gamma_{ij,k}$, where the Einstein's notation is adopted. Finally we arrive at the geodesic equations of $\nabla^e$,
\begin{equation}
\label{exp&mixtGeod}
\left\{\begin{array}{ll}
\ddot{\mu}(t)-\frac{4\, \dot{\mu}(t)\,\dot{\sigma}(t)}{\sigma(t)} &=0\\
\ddot{\sigma}(t)-\frac{3\,\dot{\sigma}^2(t)}{\sigma(t)} &=0
\end{array}\right.\, .
\end{equation}

At this point we can look for the $\nabla^e$-geodesic $\sigma^e(t)$ from $(\mu_1,\sigma_1)$ to $(\mu_2,\sigma_2)$. By assuming $0<\sigma_2<\sigma_1$, we obtain the following expression,
\begin{equation}\label{egeod}
\sigma^e(t)=\left(\frac{\mu_1\sigma_2^2(1-t)+\mu_2\sigma_1^2\,t}{\sigma_2^2+(\sigma_1^2-\sigma_2^2)t},\frac{\sigma_1\,\sigma_2}{\sqrt{\sigma_2^2+(\sigma_1^2-\sigma_2^2)t}}\right)
\end{equation}
and then
\begin{equation}
\label{dotegeod}
\dot{\sigma}^e(t)=\frac{\sigma_1\,\sigma_2}{(\sigma_2^2+(\sigma_1^2-\sigma_2^2)t)^2}\left((\mu_2-\mu_1)\sigma_1\,\sigma_2,\frac{\sigma_2^2-\sigma_1^2}{\sqrt{\sigma_2^2+(\sigma_1^2-\sigma_2^2)t}}\right)\ .
\end{equation}

Moreover, by requiring the ending points to be $(\mu_1,\sigma_1)$ and $\sigma^e(t)$, the $1$-parameter $\nabla^e$-geodesics are then given by
\begin{align}\label{etgeod}
&\sigma^e_t(s)=\left(\frac{\mu_2\sigma_1^2 t s+\mu_1\sigma_2^2(1-ts)}{\sigma_2^2+(\sigma_1^2-\sigma_2^2)ts},\frac{\sigma_1\sigma_2}{\sqrt{\sigma_2^2+(\sigma_1^2-\sigma_2^2)ts}}\right),\qquad s,t\in[0,1] . 
\end{align}
From Eq. \eqref{etgeod} we have that
\begin{equation}
\label{dotet1}
\dot{\sigma}^e_t(1)=\frac{\sigma_1\,\sigma_2\, t}{(\sigma_2^2+(\sigma_1^2-\sigma_2^2)t)^2}\left((\mu_2-\mu_1)\sigma_1\,\sigma_2,\frac{\sigma_2^2-\sigma_1^2}{\sqrt{\sigma_2^2+(\sigma_1^2-\sigma_2^2)t}}\right)\ .
\end{equation}
The last element we need for evaluating the integral \eqref{AyCDvector} is the metric tensor $\metric^{\Fi}$ computed along the exponential geodesic $\sigma^e(t)$. Then, we have that
$$
\metric(\sigma^e(t))=\left(\begin{array}{cc}
\frac{\sigma_2^2+(\sigma_1^2-\sigma_2^2)t}{\sigma_1^2\sigma_2^2}& 0\\
0& 2\, \frac{\sigma_2^2+(\sigma_1^2-\sigma_2^2)t}{\sigma_1^2\sigma_2^2}
\end{array}\right)
$$
By performing the integral \eqref{AyCDvector} we finally arrive at
$$
\Div(p_1,p_2)=\frac{(\mu_1-\mu_2)^2}{2 \sigma_1^2}+\frac{\sigma_2^2}{2 \sigma_1^2}+\ln\left(\frac{\sigma_1}{\sigma_2}\right)-1\ ,
$$
which proves that $\Div(p_1,p_2)$ is the K-L divergence between two standard normal distributions.

\vspace{.3cm}

Let us now address our investigation to the tangent dynamics described by Eq. \eqref{gradientflowsol}. In the particular case under consideration, the computation of $\grad\,\Div_{p_1}$ leads to the following expression
\begin{equation}
\label{gradnormal}
\grad\,\Div_{p_1}= \frac{(\mu(t)-\mu_1)\,\sigma^2(t)}{\sigma_1^2}\,\partial_1+ \frac{\sigma(t)}{2}\left(\frac{\sigma^2(t)}{\sigma^2_1}-1\right)\, \partial_2,
\end{equation}
where $\partial_1=\frac{\partial}{\partial\mu}$ and $\partial_2=\frac{\partial}{\partial\sigma}$. Consequently, the tangent dynamics \eqref{gradientflowsol} of the model \eqref{Gaussianmono} is described by the following ordinary differential equations,
\begin{equation}
\label{tangentdynGauss}
\left\{\begin{array}{l}
\dot{\mu}=\frac{\sigma^2(t)}{t\,\sigma_1^2}\left(\mu(t)-\mu_1\right),\quad \mu(0)=\mu_1,\quad t\in (0,1]\\
\\
\dot{\sigma}=\frac{\sigma(t)}{2\,t}\left(\frac{\sigma^2(t)}{\sigma_1^2}-1\right),\quad \sigma(0)=\sigma_1,\quad t\in (0,1]
\end{array}\right.\ .
\end{equation}
This system can be easily integrated. By assuming $\sigma_1>\sigma_2$ and recalling the positivity of $\sigma(t)$ we obtain
\begin{equation}
\label{tangentdynGausssol}
\sigma^e(t)=(\mu(t),\sigma(t))=\left\{\begin{array}{ll}
\mu(t)=\frac{\mu_1\,\sigma_2^2-t\,(\mu_1\,\sigma_2^2-\mu_2\,\sigma_1^2)}{\sigma_2^2+(\sigma_1^2-\sigma_2^2)\,t},&\quad t\in[0,1]\\
\\
\sigma(t)=\frac{\sigma_1\,\sigma_2}{\sqrt{\sigma_2^2+(\sigma_1^2-\sigma_2^2)\,t}},&\quad t\in[0,1]
\end{array}\right.\ .
\end{equation}
A slight comparison between Eq. \eqref{egeod} and Eq. \eqref{tangentdynGausssol} immediately leads to the conclusion that the latter curve is the $\nabla^e$-geodesic between $p_1$ and $p_2$.  However, we can rely on the $\nabla^e$-affine coordinates $\btheta=(\theta^1,\theta^2)$, where
$$
\theta^1=\frac{\mu}{\sigma^2},\quad \theta^2=\frac{1}{2\,\sigma^2}\ 
$$
and then we have that \eqref{tangentdynGausssol} reduces to the straight line from $p_1$ to $p_2$,
\begin{equation}\label{egeodaffine}
\btheta(t)=\left\{\begin{array}{ll}
\theta^1(t)=\theta^1_{p_1}+t\,(\theta^1_{p_2}-\theta^1_{p_1}),&\quad t\in[0,1]\\
\\
\theta^2(t)=\theta^2_{p_1}+t\,(\theta^2_{p_2}-\theta^2_{p_1}),&\quad t\in[0,1]\ ,
\end{array}\right.
\end{equation}
where $\theta^i_{p_1}$ and $\theta^i_{p_2}$ are the $\nabla^e$-affine coordinates of points $p_1$ and $p_2$, respectively.

In \cite{Obata92}, the authors proved that the Uhlenbeck-Ornstein (UO) process characterized by the Gaussian distribution
\begin{equation}
\label{UOprocess}
\mu=\mu_0+v\,(s-s_0),\quad \sigma^2=2\,D\,(s-s_0)
\end{equation}
is a geodesic with respect to $\nabla^e$-connection. Here, $\mu_0$ is the starting point at $s=s_0$ and $v$, $D$ are usually referred as the drift coefficient and the diffusion coefficient, respectively. In addition, the authors also supplied a universal relation between the "Newtonian" time $s$ and the affine time $t$ which is given by \cite{Obata92},
$
s-s_0=\frac{b}{t+a}\ ,
$
for suitable constants $a,b\in\RR$. In order to prove that the $\nabla$-geodesic solution of the tangent dynamics \eqref{tangentdynGauss} can be seen as a UO process, we can consider a rescaling of the time $t$ such that
$$
\theta^i(t)=\theta^i(0)+\frac{\tau}{t+\tau}\left(\theta^i_q-\theta^i(0)\right)
$$
and we assume that $q$ is the uniform distribution, i.e. $\theta^2_q=0$. Therefore, by setting
$$
\mu_0=\frac{\theta^1(0)-\theta^1_q}{2\,\theta^2(0)},\quad v=\frac{\theta^1(0)}{2\,\tau\,\theta^2(0)},\quad D=\frac{1}{4\,\tau\,\theta^2(0)},
$$
we obtain
$$
\mu(t)=\mu_0+v(t+\tau),\quad \sigma^2(t)=2\,D(t+\tau)
$$
which is the UO process as it is figured out in \cite{Fujiwara95}, up to some constants.

\section{Conclusions}\label{Conclusions}

Riemannian geometry enters the realm of mechanics in connection with the inertial quality of mass and principles of classical mechanics have shown that the inertia of mass is characterized by the kinetic energy  \cite{Lanczos}. Starting from this fundamental statement, in this article we established a relationship between Information Geometry (IG) \cite{Amari00} and classical mechanics \cite{Arnold} by means of the divergence function $\Div$ recently introduced by Ay and Amari in \cite{Ay15}. Given a statistical manifold $(\Ma,\metric,\nabla,\nabla^*)$, we have firstly considered the energy of an arbitrary curve $\gamma$ proposed in \cite{Amari00}. Then, we proved that such a general energy reduces to the function $\Div_{\gamma}$, which consequently is interpreted as the time integral of twice the kinetic energy \eqref{K*} in the ambient space $(\Ma,\metric,\nabla^*)$ and the time integral of twice the kinetic energy \eqref{K} in the ambient space $(\Ma,\metric,\nabla)$. In particular, by focusing our investigation on dually flat manifolds, we applied the Lagrangian formalism to both the Lagrangian functions, $\Lagrange$ given by Eq. \eqref{LagFlat} and $\Lagrange^*$ given by Eq. \eqref{Lag*}. In one case, the solution of the Euler-Lagrange equations is an unparametrized $\nabla$-geodesic; while in the other case, the Euler-Lagrange equations are solved by an unparametrized $\nabla^*$-geodesic. These results allowed us to show that both functions, the canonical divergence $\Div$ and its dual function $\Div^*$, coincide with the Hamilton principal function of $\Lagrange$ and the Hamilton principal function of $\Lagrange^*$, respectively.  

By means of these Hamilton principal functions, we moved to the Hamiltonian description of the dynamics in a general dually flat manifold. Then, we succeeded to describe the Hamiltonian equations of dynamics in term of the gradient of $\Div$ as well as in terms of the gradient of $\Div^*$. In particular, we showed that the tangent dynamics is a gradient system. According to the particular form of the Hamiltonian function \eqref{Hamiltonian}, we proved Theorem \ref{tangentdynamicsThm} which claims that the solution of the tangent dynamics is the $\nabla$-geodesic.

Finally, we applied Theorem \ref{tangentdynamicsThm} to the standard Gaussian distribution \eqref{Gaussianmono} and proved that  the canonical divergence $\Div$ is the Kullback-Leibler divergence. In addition we gave an interpretation of the tangent dynamics \eqref{gradientflowsol} in terms of the Uhlenbeck-Ornstein process  which describes the probability that a free particle in Brownian motion has a given velocity $v$ after a time $t$.

The general case is still not properly understood. More precisely, given a general statistical manifold $(\Ma,\metric,\nabla,\nabla^*)$ the geometric theory put forward in \cite{Felice18} would suggest to use the following action functionals
\begin{align*}
&\Lag(\gamma)=\int_0^1\ \langle\dot{\gamma}(t),\Pi_t (p)\rangle_{\gamma(t)}\ \total t, \qquad \Pi_t(p)=\paralleltransport_{\sigma_t^*}\dot{\sigma}_t(0)\ ,\\
& \Lag^*(\gamma)=\int_0^1\ \langle\dot{\gamma}(t),\Pi^*_t (p)\rangle_{\gamma(t)}\ \total t, \qquad \Pi^*_t(p)=\paralleltransport^*_{\sigma_t}\dot{\sigma}^*_t(0)\ ,
\end{align*}
where $\sigma_t$ and $\sigma_t^*$ are $\nabla$ and $\nabla^*$ geodesic, respectively, connecting $p$ to $\gamma(t)$ and $\paralleltransport$, $\paralleltransport^*$ denote the $\nabla$ and $\nabla^*$ parallel transports, respectively.
Here, the gradient flow of surfaces of constant $\Lag$ is obtained by projecting $\Pi_t(p)$ along the $\nabla^*$-geodesic from $p$ to $q$. Whereas, the gradient flow of surfaces of constant $\Lag^*$ is obtained by projecting $\Pi^*_t(p)$ along the $\nabla$-geodesic from $p$ to $q$. However, it remains unclear how to use the Lagrangian formalism in this context.

\bibliographystyle{plain}
\bibliography{Biblio}

\appendix

\section{Information Geometry}\label{Appendix}

In what follows, we present the main information geometric
concepts of relevance to our work. For a more detailed
description, we refer to Refs. \cite{Amari00} and \cite{Ay17}.

A {\it statistical manifold} is the quadruple $(\Ma,\metric,\nabla,\nabla^*)$, where $\metric$ is a Riemannian metric tensor and $\nabla$ and $\nabla^*$ are torsion free linear connections on $\tangent\Ma$ coupled by Eq. \eqref{dualstructure} \cite{Ay17}.

To the affine connections $\nabla$ and $\nabla^*$ we can associate two curvature tensors, 
\begin{eqnarray}
&&\Riemann(X,Y)Z=\nabla_X\nabla_Y Z-\nabla_Y\nabla_X Z-\nabla_{[X,Y]}Z,\label{curvature}\\
&&\Riemann^*(X,Y)Z=\nabla^*_X\nabla^*_Y Z-\nabla^*_Y\nabla^*_X Z-\nabla^*_{[X,Y]}Z\ . \label{curvature*}
\end{eqnarray}
where $X,Y,Z\in\Tau(\Ma)$ and $[X,Y]=XY-YX$ is the Lie bracket of $X$ and $Y$.

Given the metric structure on $\Ma$, we can also consider the Riemann curvature tensors of $\nabla$ and $\nabla^*$ that are defined as follows
\begin{align}
\label{RiemC}
&\RC(X,Y,Z,W):=\metric\left(\Riemann(XY)Z,W\right)\\
&
\label{RC*}
\RC^*(X,Y,Z,W):=\metric\left(\Riemann^*(XY)Z,W\right)\ .
\end{align}
The relation between $\RC$ and $\RC^*$ is given by \cite{Lauritzen87} 
\begin{equation}\label{alter}
\RC(X,Y,Z,W)=-\RC^*(X,Y,W,Z)\ .
\end{equation}

For a general statistical manifold $(\Ma,\metric,\nabla,\nabla^*)$ it is not true that
\begin{equation}
\label{ConjSymm}
\RC(X,Y,Z,W)=-\RC(X,Y,W,Z)\ .
\end{equation}
This condition identifies a class of statistical manifolds called {\it conjugate symmetric} \cite{Lauritzen87}. Very remarkably, for a conjugate symmetric statistical manifold the following property holds true:
\begin{equation}
\label{ConjSymmRC}
\RC(X,Y,Z,W)=\RC^*(X,Y,Z,W), \quad \forall X,Y,Z,W \in \Tau(\Ma)\ .
\end{equation}
Finally, we call $(\Ma,\metric,\nabla,\nabla^*)$ a {\it dually flat} statistical manifold if it is conjugate symmetric and $\RC\equiv 0$.

\vspace{.2cm}

Dually flat manifolds are widely studied by Amari and Nagaoka \cite{Amari00}. In particular, given a dually flat manifold $(\Ma,\metric,\nabla,\nabla^*)$, there exist affine coordinate systems $\theta=\{\theta^1,\ldots,\theta^n\}$ with respect to $\nabla$ and $\eta=\{\eta_1,\ldots,\eta_n\}$ with respect to $\nabla^*$ such that
\begin{equation}
\label{biorthogonal}
\metric\left(\frac{\partial}{\partial \theta^i},\frac{\partial}{\partial \eta_j}\right)=\delta^i_j,
\end{equation}
where $\delta^i_j$ denotes the delta-Dirac function. These coordinate systems are called {\it dual coordinate systems}. In this case, there exist two functions $\Psi,\varPhi:\Ma\rightarrow\RR$ such that $\total \Psi=\eta_i\total\theta^i$ and $\total \varPhi=\theta_i\total\eta^i$. The correspondence between $\theta$ and $\eta$ is given in terms of the functions $\Psi$ and $\varPhi$ by means of the following {\it Legendre transformation}
\begin{equation}
\label{Legendre}
\varPhi+\Psi=\theta^i\eta_i,
\end{equation}
where the summation over $i=1,\ldots,n$ is intended. Through $\varPhi$ and $\Psi$ a function $\D:\Ma\times\Ma\rightarrow\RR$ can be defined by
\begin{equation}
\label{CanonicalDivergence}
\D(p,q):= \varPhi(p)+\Psi(q)-\eta_i(p)\theta^i(q),\quad \forall p,q\in\Ma\ ,
\end{equation}
and it is called the {\it canonical divergence} on $\Ma$ of Bregman type \cite{Amari00}.

\subsection{Canonical Divergence within Information Geometry}

In this section we describe the canonical divergence introduced in  \cite{Ay15} by Ay and Amari. It is given by using the geodesic integration of the inverse exponential map. This one is interpreted as a {\it difference} vector that translates $q$ to $p$ for all $q,p$ suitably close in $\Ma$. 

\begin{figure}[h!]\label{differencevect}
\centering
\includegraphics[scale=1]{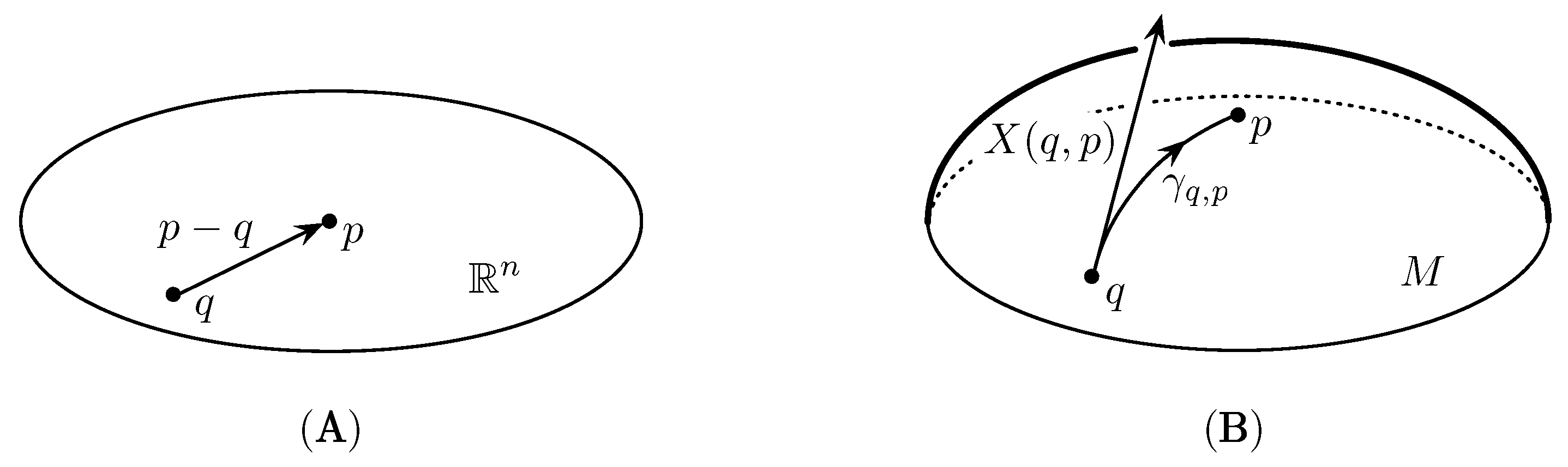}
\caption{On the left, (A) illustrates the difference vector $p-q$ in the linear vector space $\RR^n$; whereas, in (B) we can see the difference vector $\nihat(q,p)=\dot{\gamma}_{q,p}(0)$ in $\Ma$ as the inverse of the exponential map at $q$ (This Figure comes from \cite{Ay15}).}
\end{figure}

To be more precise, the inverse exponential map provides a generalization to $\Ma$ of the notion of difference vector of the linear vector space. In detail, let $p,q\in\RR^n$, the difference between $p$ and $q$ is given by the vector $p-q$ pointing to $p$ (see side (A) of Fig. \ref{differencevect}). Then, the difference between $p$ and $q$ in $\Ma$ is supplied by the exponential map of the connection $\nabla$. In particular, assuming that $p\in \mathrm{U}_q$ and $\mathrm{U}_q\subset \Ma$ is a $\nabla$-geodesic neighborhood of $q$, the difference vector from $q$ to $p$ is defined as (see (B) of Fig. \ref{differencevect})
\begin{equation}
\label{nihat}
\nihat_q(p):=\nihat(q,p):=\exp_q^{-1}(p)=\dot{\gamma}_{q,p}(0)\ ,
\end{equation}
where $\gamma_{q,p}$ is the $\nabla$-geodesic from $q$ to $p$ laying in $\mathrm{U}_q$. Clearly, by fixing $p\in\Ma$ and letting $q$ vary in $\Ma$, we obtain a vector field $\nihat(\cdot,p)$ whenever a $\nabla$-geodesic from $q$ to $p$ exists. From here on, we equally use both the notations, $\nihat(q,p)$ and $\nihat_q(p)$, for representing the difference vector from $q$ to $p$. {Obviously, $\nihat_p(q)$ denotes the difference vector from $p$ to $q$.}

Therefore, the divergence proposed by Ay and Amari in \cite{Ay15} is defined as the path integral
\begin{equation}
\label{canonicaldiv}
\Div_{\gamma}(p,q):=\int_0^1 \langle \nihat_t(p),\dot{\gamma}(t)\rangle_{\gamma(t)}\ dt\ ,
\end{equation}
where $\gamma$ is the $\nabla$-geodesic from $q$ to $p$ and $\langle\cdot,\cdot\rangle_{\gamma(t)}$ denotes the inner product with respect to $\metric$ evaluated at $\gamma(t)$. In Eq. \eqref{canonicaldiv}, $\nihat_t(p)$ is the vector field along $\gamma(t)$ given by Eq. \eqref{nihat} as follows,
\begin{equation}
\label{nihatvstime}
\nihat_t(p)=\nihat(\gamma(t),p)=\exp_{\gamma(t)}^{-1}(p)\ . 
\end{equation}
{After elementary computation Eq. \eqref{canonicaldiv} reduces to \cite{Ay15},
\begin{equation}
\label{Aydivergence}
\Div_{\gamma}(p,q)=\int_0^1 \ t \|\dot{\gamma}_{p,q}(t)\|^2\ dt\ ,
\end{equation}
where $\gamma_{p,q}(t)$ is the $\nabla$-geodesic from $p$ to $q$.
If we consider definition \eqref{canonicaldiv} for general path $\gamma$ then $\Div_{\gamma}(p,q)$ will be depending on $\gamma$. On the contrary, if the vector field $\nihat_t(p)$ is integrable, then $\Div_{\gamma}(p,q)=: \Div(p,q)$ turns out to be independent of the path from $q$ to $p$. }

\subsection{Gradient fields in dually flat manifolds}\label{Agradientfields}

Consider a dually flat manifold $(\Ma,\metric,\nabla,\nabla^*)$. In order to prove that \eqref{differencevect} is a gradient vector field, we recall that the canonical divergence $\Div_p(q)$ assumes the following form,
$$
\Div_p(q)=\int_0^1\ \langle\dot{\sigma}_t(1),\dot{\sigma}(t)\rangle_{\sigma(t)}\ \total t,
$$
where $\sigma$ is the $\nabla$-geodesic from $p$ to $q$ and $\sigma_t$ is the $\nabla$-geodesic from $p$ to $\sigma(t)$. Let us observe that $\nihat_t(p)=-\dot{\sigma}_t(1)$. We shall show that 
$
\dot{\sigma}_t(1)=\grad_{t}\Div_p\ .
$
To accomplish this aim, we firstly need the following result. This is well-known in literature (see Ref. \cite{Amari00}). However, here we propose an alternative derivation.

\begin{pro}
\label{PerpendicularProp}
Let $(\Ma,\metric,\nabla,\nabla^*)$ be a dually flat statistical manifold. Then each $\nabla$-geodesic from $p$ is perpendicular to the level-hypersurfaces of the function $\Div_p(\cdot)$.
\end{pro}
\noindent {\bf Proof.}\; Let $\Sigma(\tau,t)\ (-\varepsilon<\tau<\varepsilon, \ 0\leq t\leq 1)$ be a variation of $\nabla$-geodesics such that
$$
\left\{\begin{array}{l}
\Sigma(0,t)=\sigma(t)\\
\Sigma(\tau,0)=p \; \forall \tau\\
\langle V, T\rangle(1)=0
\end{array}\right.
$$
where $V:=\partial_{\tau}\Sigma(0,t)$ is the $\nabla$-Jacobi vector field and $T:=\partial_t \Sigma(\tau,t)$ is the velocity vector along $\sigma(t)$. Consider now a $\nabla$-geodesic $\sigma_{\tau,t}(s)$ such that $\sigma_{\tau,t}(0)=p$ and $\sigma_{\tau,t}(1)=\Sigma(\tau,t)$. By defining $S:=\partial_s \sigma_{\tau,t}(1)$, we immediately get the proportionality relation $S=t\ T$. In this setting, we have then
\begin{eqnarray*}
\langle\grad\ \Div_p(\cdot),\partial_{\tau}\Sigma(0,1)\rangle&=&\left.\frac{\total}{\total\tau}\right|_{\tau=0} \Div_p(\Sigma(\tau,1))=\int_0^1\ V\langle T,S\rangle\ \total t \\
&=& \int_0^1 \left(\langle \nabla^*_V T,S\rangle+\langle T,\nabla_V S\rangle\right)\total t\\
&=& \int_0^1 \left(\langle \nabla^*_T V,S\rangle+\langle T,\nabla_V S\rangle\right)\total t\\
&=& \int_0^1 \left(T\langle V,S\rangle-\langle V,\nabla_T S\rangle+\langle T,\nabla_V S\rangle\right)\total t,
\end{eqnarray*}
where we used the torsion freeness of $\nabla^*$ and the dual property \eqref{dualstructure} of $\nabla$ and $\nabla^*$. By recalling that $S=t\ T$ we can observe that
\begin{align*}
& \langle V,\nabla_T t T\rangle= \langle V,T+\nabla_T  T\rangle= \langle V,T\rangle\\
& \langle T,\nabla_V t T\rangle= \langle T,0+\nabla_V T\rangle,
\end{align*}
where the first line is obtained because $\sigma$ is $\nabla$-geodesic and the second line follows by taking the derivative with respect to $\tau$ as indicated by the definition of $V$. Therefore, we have that
$$
\langle\grad\ \Div_p(\cdot),\partial_{\tau}\Sigma(0,1)\rangle=\left[\langle V,T\rangle\right]^1_0-\int_0^1\ \langle V,T\rangle\ \total t+\int_0^1\ t\langle T,\nabla_T V\rangle\ \total t\ ,
$$
where we used the torsion freeness of $\nabla$. 

From classical Riemannian Geometry we know that $V$ is a Jacobi field of a $\nabla$-geodesic variation if and only if \cite{Lee97}
\begin{equation}
\label{Jacobi}
D^2_t\ V+\Riemann(V,\dot{\sigma})\dot{\sigma}=0\ ,
\end{equation}
where $\Riemann$ denotes the curvature tensor of $\nabla$ given by Eq. \eqref{curvature}. Since $\Ma$ is $\nabla$-flat, Eq. \eqref{Jacobi} reduces to $D^2_t\ V=0$. This implies that, if we set $W=\nabla_T V$ then $W$ is $\nabla$-parallel along $\sigma(t)$, i.e. $D_t W=0$.

Let us now assume that $V(0)=0$ and $D_t V(0)=W(0)$. Because of $\nabla$-flatness the solution of $D_t^2 V=0$ is given by
$
V(t)=t\ W(t)
$.
Therefore, we have that
\begin{eqnarray*}
\langle\grad\ \Div_p(\cdot),\partial_{\tau}\Sigma(0,1)\rangle&=&\left[\langle V,T\rangle\right]^1_0-\int_0^1\ \langle V,T\rangle\ \total t+\int_0^1\ t\langle T,\nabla_T V\rangle\ \total t\ \\
&=& -\int_0^1\left( \langle t\ W,T\rangle-t\langle T,\nabla_T\ t W\rangle\right)\total t\\
&=& -\int_0^1\left(t \langle  W,T\rangle-t\langle T,\nabla_T  W\rangle-t\langle  W,T\rangle\right)\total t \\
&=& 0\ .
\end{eqnarray*}
This proves the desired result. \hfill $\square$

\begin{remark}\label{RemPerpendicular}
The dual version of Proposition \ref{PerpendicularProp} holds true, as well. More precisely, given a dually flat statistical manifold $(\Ma,\metric,\nabla,\nabla^*)$ the dual function of the canonical divergence reads as follows,
$$
\Div^*_p(q)=\int_0^1\ \langle\dot{\sigma}^*_t(1),\dot{\sigma}^*(t)\rangle_{\sigma^*(t)}\ \total t,
$$
where $\sigma^*$ is the $\nabla^*$-geodesic from $p$ to $q$ and $\sigma^*_t$ is the $\nabla^*$-geodesic from $p$ to $\sigma^*(t)$.
Then, each $\nabla^*$-geodesic from $p$ is perpendicular to the level-hypersurfaces of the function $\Div^*_p(\cdot)$.

\vspace{.2cm}

Actually, this result has been obtained by Henmi and Kobayashi in a context more general than dually flatness \cite{Kobayashi00}. They have considered the claim of Proposition \ref{PerpendicularProp} under the conditions
$$
\nabla\RC=0,\quad \RC(X,Y,Y,Y)=0\quad \forall\ X,Y\in\Tau(M),
$$
which are obviously implied by dually flatness.
\end{remark}

\begin{theorem}\label{Thgrad}
Let $(\Ma,\metric,\nabla,\nabla^*)$ be a dually flat manifold. Then, we have that
\begin{equation}
\label{gradientdifferencevector}
\dot{\sigma}_t(1)=\grad_{\sigma(t)}\Div_p\ ,
\end{equation}
where $\sigma(t)$ is the $\nabla$-geodesic between $p,q\in\Ma$ and $\sigma_t(s)$ is the $\nabla$-geodesic from $p$ to $\gamma(t)$.
\end{theorem}
\noindent{\bf Proof.}\, Consider a $\nabla$-geodesic $\sigma(t)$ such that $\sigma(0)=p$ and $\sigma(1)=q$. Then, for all $t\in[0,1]$ consider the $\nabla$-geodesic $\sigma_t(s)$ such that $\sigma_t(0)=p$ and $\sigma_t(1)=\sigma(t)$. We have that
$$
\Div_p(q)=\int_0^1\ \langle\dot{\sigma}_t(1),\dot{\sigma}(t)\rangle_{\sigma(t)}\ \total t\ .
$$
Consider now a variation $q_{\tau}=\sigma_1(1+\tau)$ of the endpoint $q$. By considering the $\nabla$-geodesic variation $\Sigma(\tau,t)$ such that $\Sigma(\tau,0)=p$ and $\Sigma(\tau,1)=q_{\tau}$ we can apply methods of proposition \eqref{PerpendicularProp} and then we have that
\begin{eqnarray*}
\langle \grad\ \Div_p(\cdot),\left.\frac{\total}{\total \tau}\right|_{\tau=0}q_{\tau}\rangle &=& \left.\frac{\total}{\total \tau}\right|_{\tau=0} \Div_p(\cdot)\\
&=& \langle \dot{\sigma}_1(1),\left.\frac{\total}{\total \tau}\right|_{\tau=0}q_{\tau}\rangle\ .
\end{eqnarray*}
From Proposition \eqref{PerpendicularProp} we know that $\nabla$-geodesics are perpendicular to level-hypersurfaces $\Div_p(\cdot)$. This implies
$$
\grad\ \Div_p(q)=\dot{\sigma}_1(1)\ .
$$
In particular, we have that $\grad_t\Div_p=\grad\ \Div_p(\sigma(t))=\dot{\sigma}_t(1)$. \hfill $\qed$

\end{document}